\documentclass[prapplied,amsmath,amssymb, aps,graphicx,twocolumn,showpacs]{revtex4-2}
\usepackage{graphicx}
\usepackage{physics}
\usepackage{color}

\begin{document}
\newcommand{\dif}{\mathrm{d}}

\title{Improved superconducting qubit state readout by path interference}

\author{Zhiling Wang}
\altaffiliation{These two authors contributed equally to this work.}

\author{Zenghui Bao}
\altaffiliation{These two authors contributed equally to this work.}

\author{Yukai Wu}

\author{Yan Li}

\author{Cheng Ma}

\author{Tianqi Cai}

\author{Yipu Song}

\author{Hongyi Zhang}
\email{hyzhang2016@tsinghua.edu.cn}

\author{Luming Duan}
\email{lmduan@tsinghua.edu.cn}

\affiliation{Center for Quantum Information, Institute for Interdisciplinary Information Sciences, Tsinghua University, Beijing 100084, PR China}

\date{\today}

\begin{abstract}
High fidelity single shot qubit state readout is essential for many quantum information processing protocols. In superconducting quantum circuit, the qubit state is usually determined by detecting the dispersive frequency shift of a microwave cavity from either transmission or reflection. In this paper, we demonstrate the use of constructive interference between the transmitted and reflected signal to optimize the qubit state readout, with which we find a better resolved state discrimination and an improved qubit readout fidelity. As a simple and convenient approach, our scheme can be combined with other qubit readout methods based on the discrimination of cavity photon states to further improve the qubit state readout.

\end{abstract}


\maketitle

\section{\label{main:intro}Introduction}

High fidelity measurement of qubit states is an essential requirement for numerous protocols in quantum computation~\cite{DiVincenzo2000} and quantum information processing \cite{DiVincenzo2009,Wallraff13,Martinis2014,Martinis15nature} and it lies at the heart of many quantum technologies~\cite{Siddiqi2012,DiCarlo2012,Devoret2013,Wallraff2018}. 
For example, in a noisy intermediate-scale quantum computer, a quantum circuit is expected to run for a lot of trails to mitigate the possible errors \cite{Martinis19nature,IBM2019}.
A higher qubit readout fidelity would greatly improve the probability of successful trials, and thus leading to a higher computing efficiency \cite{Tannu2019,Tannu2019a}. 
Recently, determining the state of various kinds of physical qubits with high fidelity becomes feasible, even in a quantum non-demolition (QND) way \cite{Mooij2007,Mooij2010,Tarucha2019,Raha2020}. 
Fast and QND qubit state readout in a single shot is essential for a fault tolerant quantum computer, where errors have to be corrected in situ once being detected~\cite{Martinis15nature,DiCarlo15,Siddiqi16}.

In a superconducting circuit quantum electrodynamics (cQED) system, the most commonly used qubit readout scheme is based on the dispersive interaction between the qubit and the cavity field. The cavity's response to coherent state photons is dependent on the state of the qubit, which acts as an intra-cavity pointer state \cite{Schoelkopf04,Esteve09}. The qubit state can be readily determined if the two pointer states, corresponding to the qubit ground state $\ket{g}$ and the excited state $\ket{e}$, are well separated in the phase space. Ideally a probe tone with more photons, and thus larger separation of the two pointer states, is preferred to achieve a fast and high fidelity qubit state measurement \cite{Wallraff17}, but in practice a strong probe would induce notable back-action to the qubit state, which is unfavorable for a QND measurement \cite{Nori19}. Therefore, it is highly desired to make full use of photons in the cavity to effectively read out the qubit state and ultimately achieve a quantum limit measurement \cite{Girvin03,Schoelkopf2010RPM}.

For a symmetric cavity with identical out-coupling rates at both ends, one always losses half of the signal by measuring only the transmission or reflection of the cavity \cite{Wallraff2020review}. One solution is to use asymmetric cavity with the out-coupling rate of one end significantly larger than that of the other \cite{Martinis15nature,Wallraff17}. The cavity photons would mainly leak out from the end with larger out-coupling rate, and thus can be effectively collected to determine the qubit state.
In this work, we provide a complementary solution and show that it is possible to extract all the photons from a symmetric cavity through interference, thus increasing the separation of the two pointer states to reach its maxima under the given incoming photons. To demonstrate the idea, experimentally we design a sample containing five qubits with varied interference conditions for microwave photons in the readout cavities. 
With a set of hybrid couplers we are able to carry out qubit state measurement simultaneously using the output signal from the transmission path ($T$),  the reflection path ($R$) and the interference path ($T+R$). It turns out that when constructive interference condition is satisfied, the cavity response measured from $T+R$ is enhanced compared with that from $T$ or $R$. Further, qubit state discrimination and qubit readout fidelity also show clear improvement when measuring from the $T+R$ output. Finally, it is worth noting that since our method provides an effective extraction of cavity photons, all those readout schemes based on detecting qubit state dependent cavity response, including cavity driving, nonlinear qubit-cavity interaction \cite{Mottonen19,Devoret19,Schoelkopf2010,Blais10,Wallraff2020review}, etc., would benefit from the path interference method proposed in this work.

\begin{figure}[htbp]
\includegraphics[width=1\linewidth]{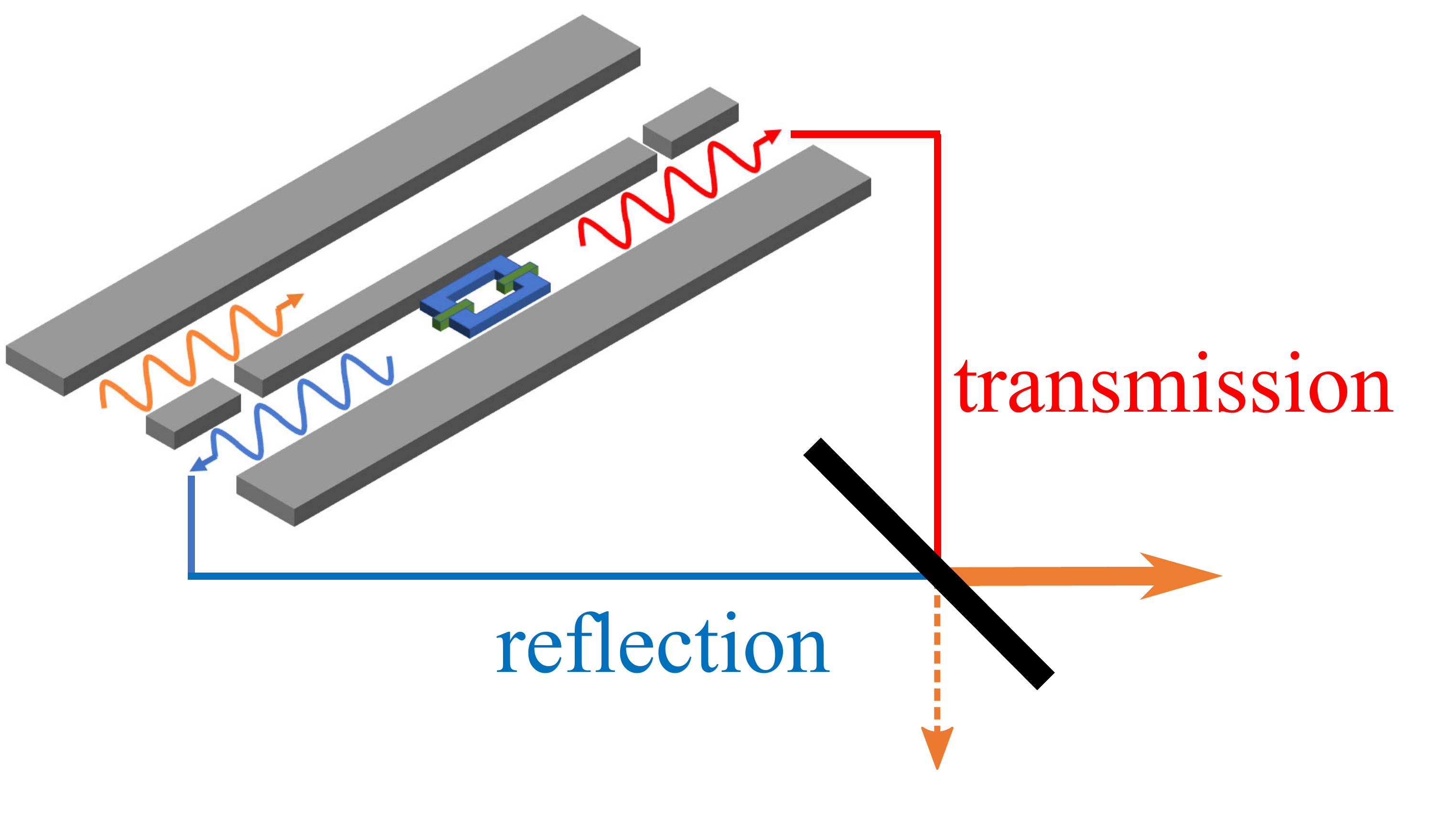}
\caption{Qubit state readout with path interference. Reflection and transmission of the readout cavity are interfered with a beam splitter to fully extract photons in the cavity.}
\label{fig_device}
\end{figure}

\section{\label{main:theory}Theory}

We consider a coplanar waveguide (CPW) cavity with symmetric out-coupling to the transmission line and dispersively coupled to a superconducting qubit, as illustrated in Fig.~\ref{fig_device}. Normally the qubit state is determined by detecting the cavity response to a probe tone  from the output $T$ or $R$. Here we consider using a beamsplitter to combine the two outputs and generate interference of the cavity field. Intuitively, the cavity photons can be effectively extracted once the constructive interference condition is met, and thus leading to a better qubit readout performance.
The Hamiltonian of the system can be written as $H=(\omega_r+\chi\sigma_z)a^\dagger a+\omega_q \sigma_z/2$, where $\omega_r$ is the bare frequency of the cavity, $\omega_q$ is the qubit transition frequency and $\chi$ is the dispersive shift due to the qubit-cavity coupling. The change of the qubit state would shift the cavity resonance through the term $\chi\sigma_z$, resulting in a change of the cavity response when probing the system at a certain frequency. For output $T$ and $R$, the distance between the two pointer states corresponding to the qubit state $\ket{g}$and $\ket{e}$ can be given as (see Supplementary Information for details)
\begin{equation}
\begin{split}
D^{T(R)}&=|\alpha_{T(R)}(\omega_d,1)-\alpha_{T(R)}(\omega_d,-1)|\\
&=\frac{4\kappa_c\chi|\alpha_{in}|}{\sqrt{(\kappa^2+4\chi^2-4(\omega_d-\omega_r)^2)^2+16\kappa^2(\omega_d-\omega_r)^2}}\\
\label{tr-dis-main}
\end{split}
\end{equation}
where $\alpha_{T(R)}(\omega_d,1(-1))$ represents intra-cavity pointer state measured from the output $T$($R$) with probing frequency $\omega_d$ and qubit state $\sigma_z = 1$ or $-1$, $\kappa$ and $\kappa_c$ are the cavity's total damping rate and the damping rate to the transmission line, respectively.

\begin{figure}[htbp]
\includegraphics[width=1\linewidth]{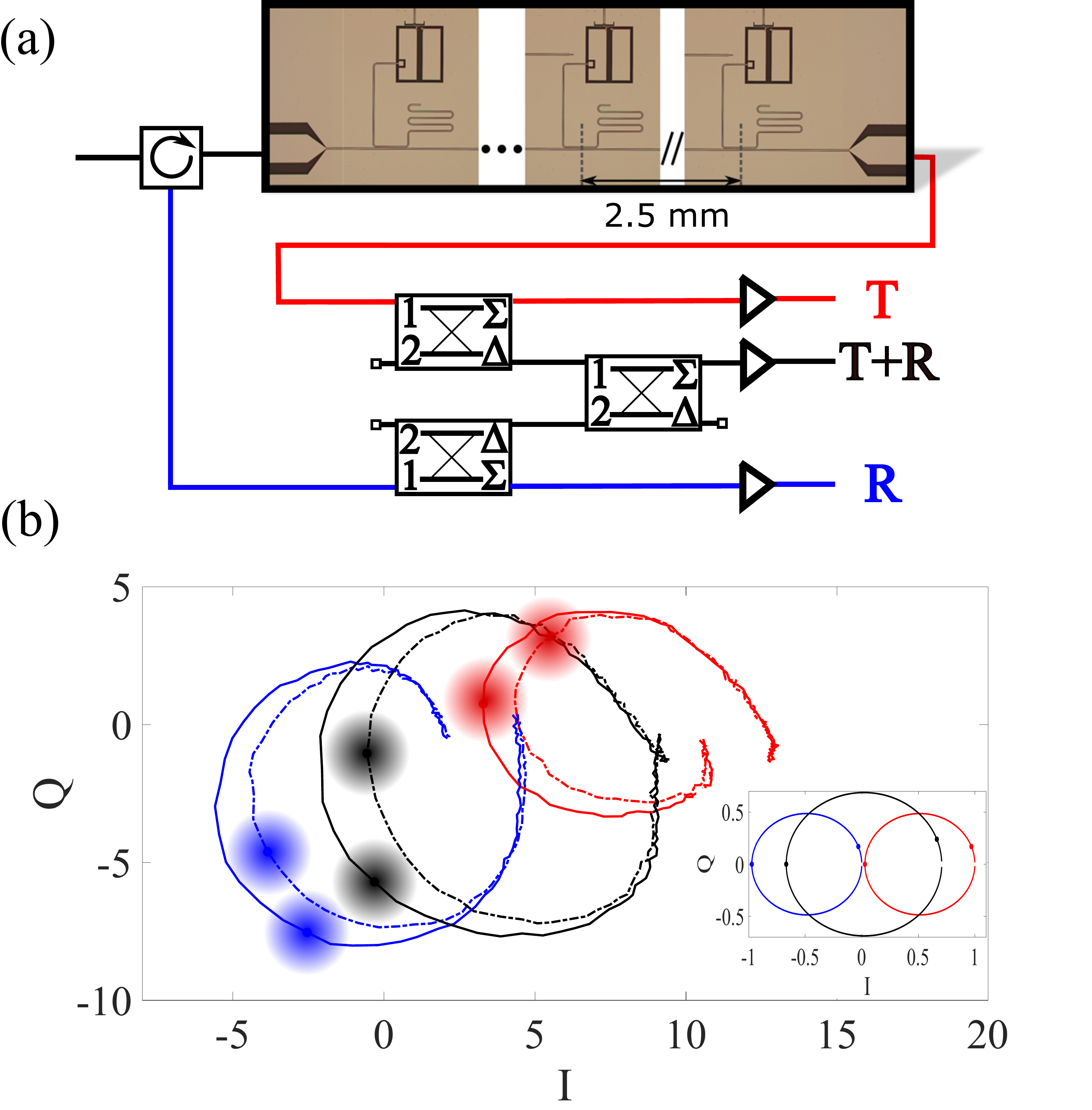}
\caption{A demonstration of the proposed scheme. (a)  An illustration of the sample and the interference circuit. The device contains five qubits, with the readout cavities equally spaced by about $2.5\,$mm along the transmission line. 
The cavity reflection or transmission is first splitted by hybrid couplers into two channels. One is sent to the corresponding amplification chain and the other is sent to a third hybrid coupler to generate interference signal. Only one interference output is sent to the amplification chain and the other one is terminated. With the help of this setup, we can realize qubit state readout through $T$, $R$ and $T+R$ simultaneously. (b) The steady state cavity responses measured from $T$ (red), $R$ (blue), $T+R$ (black) are plotted on the phase plane, when sweeping the probe frequency across the cavity resonance. The relative phase $\theta_{RT}$ for this cavity is 0.11, with which the constructive interference condition can be approximately satisfied. The solid and dashed lines are measurement results when the qubit is set to $\ket{g}$ and $\ket{e}$. The colored dots on the circles correspond to cavity responses at  probe frequencies used in the single shot measurements. The two dimensional Gaussian profiles represent the corresponding coherent states. Inset of (b) are the calculated cavity responses for $T$, $R$ and $T+R$, taking $\theta_{RT}=0$.}
\label{fig_IQ}
\end{figure}

If a beamsplitter is used to interfere output $T$ and $R$, photon state at the two output ports of the beamsplitter can be written as
\begin{equation}
\alpha_{int}^\pm(\omega_d,\sigma_z)= \frac{\alpha_{T}(\omega_d,\sigma_z) \pm e^{i\theta_{RT}}\alpha_{R}(\omega_d,\sigma_z)}{\sqrt{2}},
\label{alpha-int-main}
\end{equation}
where $\theta_{RT}$ is the relative phase between $T$ mode and $R$ mode before interference. Therefore, the distance between the two pointer states measured from the interference output can be written as
\begin{equation}
\begin{split}
D^{int}(\theta)=|\alpha_{int}(\omega_d,1)\pm \alpha_{int}(\omega_d,-1)|=\left|\frac{1\pm e^{i\theta_{rt}}}{\sqrt{2}}\right| D^{T(R)}
\end{split}
\label{eq-inter-ratio}
\end{equation}
Inset of Fig.~\ref{fig_IQ}(b) gives calculated cavity response on the phase plane when sweeping the probe frequency across the cavity resonance. The colored dots represent pointer states when the qubit is at the ground state $\ket{g}$ and the excited state $\ket{e}$ for a fixed probing frequency with a dispersive shift $\chi$. 
 From Eq.~(\ref{tr-dis-main}), distances between the two pointer states are equal for output $T$ and output $R$, which is not surprising for a symmetric cavity. It also means that if only collecting from either $T$ or $R$, one would lose half of the cavity photons, which leads to a loss of information about the qubit state up to $1/2$ \cite{Girvin03}. 
From Eq.~(\ref{eq-inter-ratio}) it can be found that once the constructive interference condition is satisfied, the distance of the cavity responses from the interference output can be maximally enlarged by $\sqrt{2}$ times than that from $T$ or $R$. Consequently the pointer state discrimination can be improved, which is preferred for the qubit state readout, as illustrated in Fig.~\ref{fig_IQ}(b).

\section{\label{main:result}Experimental results}

In the experiment, we use a sample containing five transmon qubits, each being coupled to a hanger type CPW cavity for qubit state readout. The hanger is a typical symmetric cavity with only one port coupled to the transmission line \cite{gao2008physics}. As shown in Eq.~(\ref{eq-inter-ratio}), the distance of the two pointer states from interference output depends on the relative phase $\theta_{RT}$, which experimentally can be tuned via a cryogenic phase shifter \cite{Bradley2018,Mikko2017,Mikko2020,Naaman2017}. Here to demonstrate the idea, we vary $\theta_{RT}$ by placing the readout cavity at different positions of the transmission line.
For photons from a certain cavity going through $T$ and $R$, the phase difference between the two paths depends on the position of the readout cavities on the transmission line, as illustrated in Fig. \ref{fig_IQ}(a). 
Detailed information about the sample are listed in the Supplementary Information.

We use three 180 degree microwave hybrid couplers as the beamsplitters to generate path interference \cite{Gross2018,Gross2019}. Transmission or reflection of the cavities is firstly guided to a coupler and is splitted into two channels. One channel is directly sent to the amplification chain as the output $T$ or $R$. The other channel is sent to the third coupler for interference. One of the interference output is sent to the amplification chain as the output $T+R$. It is necessary to mention that since $T$, $R$, $T+R$ are outputted from three different amplification chains, careful calibrations have to be carried out before any comparison among them. Details about the calibration work can be found in Part II of the Supplementary Information.

We first characterize the relative phase $\theta_{RT}$ for each of the readout cavities. For different output lines, the phase delays from hybrid couplers to the room temperature homodyne detector are not the same, therefore $\theta_{RT}$ cannot be directly obtained from the measured cavity responses from room temperature electronics. Instead, for a certain output line, we use cavity responses corresponding to different qubit states as reference to calibrate $\theta_{RT}$. To be more specific, we set the qubit state to $\ket{g}$ or $\ket{e}$, then record the corresponding cavity responses from $T$, $R$ and $T+R$. $\theta_{RT}$ can be deduced by comparing the cavity responses from the same output line when the qubit is in $\ket{g}$ or $\ket{e}$.  Details of this method can be found in Part III of the Supplementary Information. For the sample used in this work we have achieved various $\theta_{RT}$ from $-1.42$ to $2.60$ for the five readout cavities. In the following experiment, we use the readout cavity with $\theta_{RT} = 0.11$ to investigate the effect of constructive interference.

\begin{figure}[htbp]
\includegraphics[width=1\linewidth]{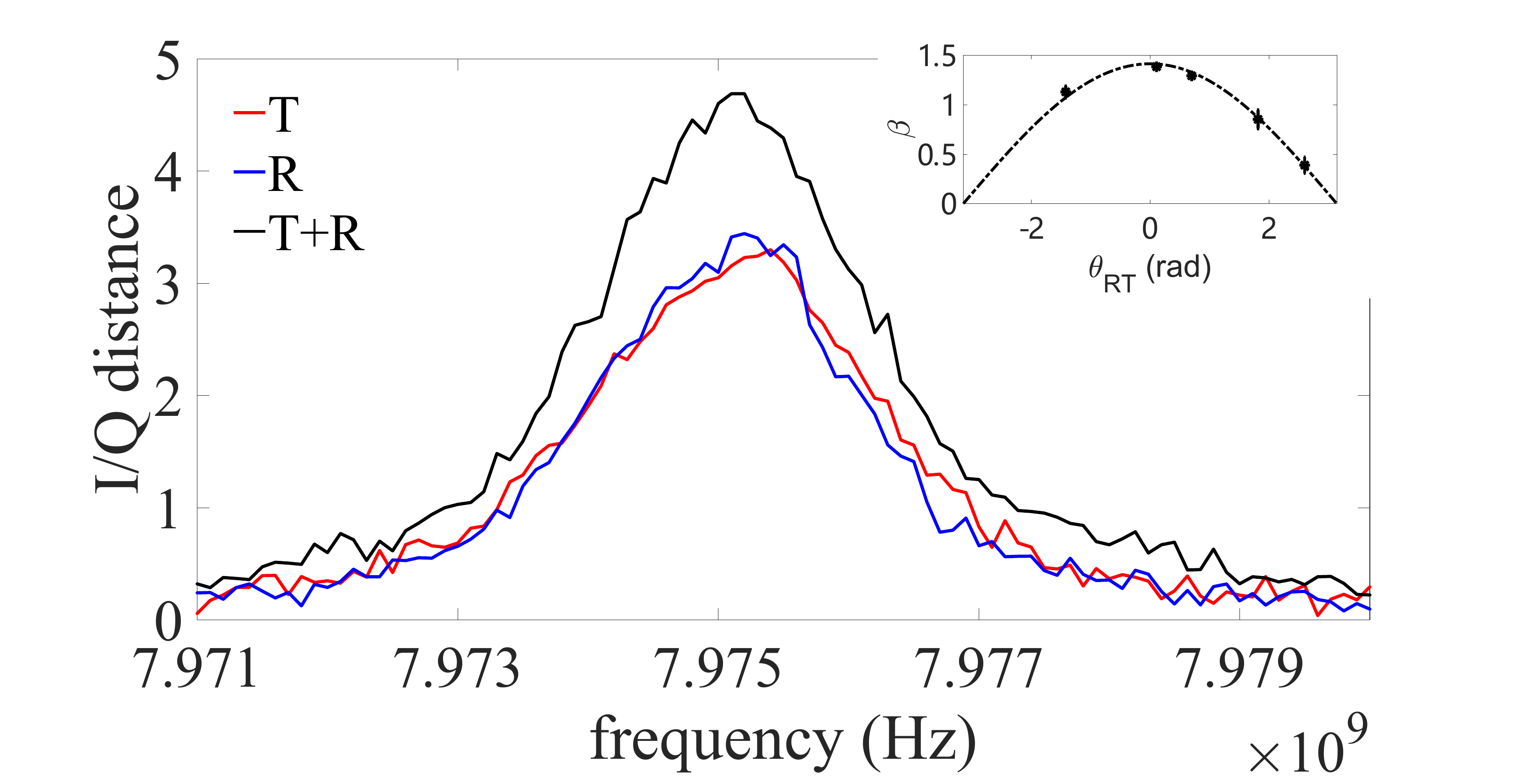}
\caption{The distance between the pointer states corresponding to the qubit states $\ket{g}$ or $\ket{e}$ on the phase plane when scanning the probe frequency across the cavity resonance. The probe frequency with the maximal distance can be used for qubit state readout, which shall yield the best signal to noise ratio. The relative phase $\theta_{RT}$ for this cavity is 0.11, which is close to the ideal constructive interference condition. The maximum value of the interference signal is about $\sqrt{2}$ times as large as that from $T$ or $R$, as predicted from the theory. The inset is the interference gain $\beta$ as a function of $\theta_{RT}$. The dashed line is the theoretical prediction and the scattered points are experimental results. The error bar of the experimental results are given by a standard deviation of the measured steady state output.}
\label{fig_IQdis}
\end{figure}

As mentioned above, qubit state readout relies on discriminating the cavity responses at a certain frequency, thus a larger distance $D(\omega_d)$ is preferred. 
In order to evaluate cavity interference effect, we sweep the probe frequency and record cavity responses from output $T$, $R$ and $T+R$ when the qubit state is in $\ket{g}$ or $\ket{e}$ \cite{Weides2015}. Fig.~\ref{fig_IQ}(b) gives such a result for the cavity with $\theta_{RT} = 0.11$, for which the constructive interference condition is almost met. Different from the theoretical result shown in inset of Fig.~\ref{fig_IQ}(b), the measured cavity response when the qubit being in $\ket{e}$ is a smaller circle on the phase plane than that when the qubit is in $\ket{g}$. This is due to the qubit relaxation during the cavity response measurement. Detailed discussion can be found in the Supplementary Information. 
From the measured cavity response in Fig.~\ref{fig_IQ}(b) we can calculate the distance between cavity responses corresponding to the qubit state $\ket{g}$ and $\ket{e}$ with varied probe frequencies, as illustrated in Fig.~\ref{fig_IQdis}. The probe frequency with the largest distance can be used for qubit state readout, which shall give the best signal to noise ratio.  
Here we introduce an enhancement factor $\beta = 2D_m^{T+R}/(D_m^{T}+D_m^{R})$ to represent the readout enhancement when measuring from $T+R$ compared with measuring $T$ or $R$, where $D_m^{T+R(T,R)}$ represents the largest distance calculated from the measured cavity response. Inset of Fig.~\ref{fig_IQdis} shows experimental results of $\beta$ with scattered plots. Theoretically one would have $\beta=\sqrt{2}\cos(\theta_{RT}/2)$ from Eq.~(\ref{eq-inter-ratio}), which is plotted as the dashed line. It can be seen that the experimental result fits well with the theoretical prediction. When the constructive interference condition is met, we obtain an enhancement factor $\beta \sim 1.4$.

\begin{figure}[htbp]
\includegraphics[width=0.9\linewidth]{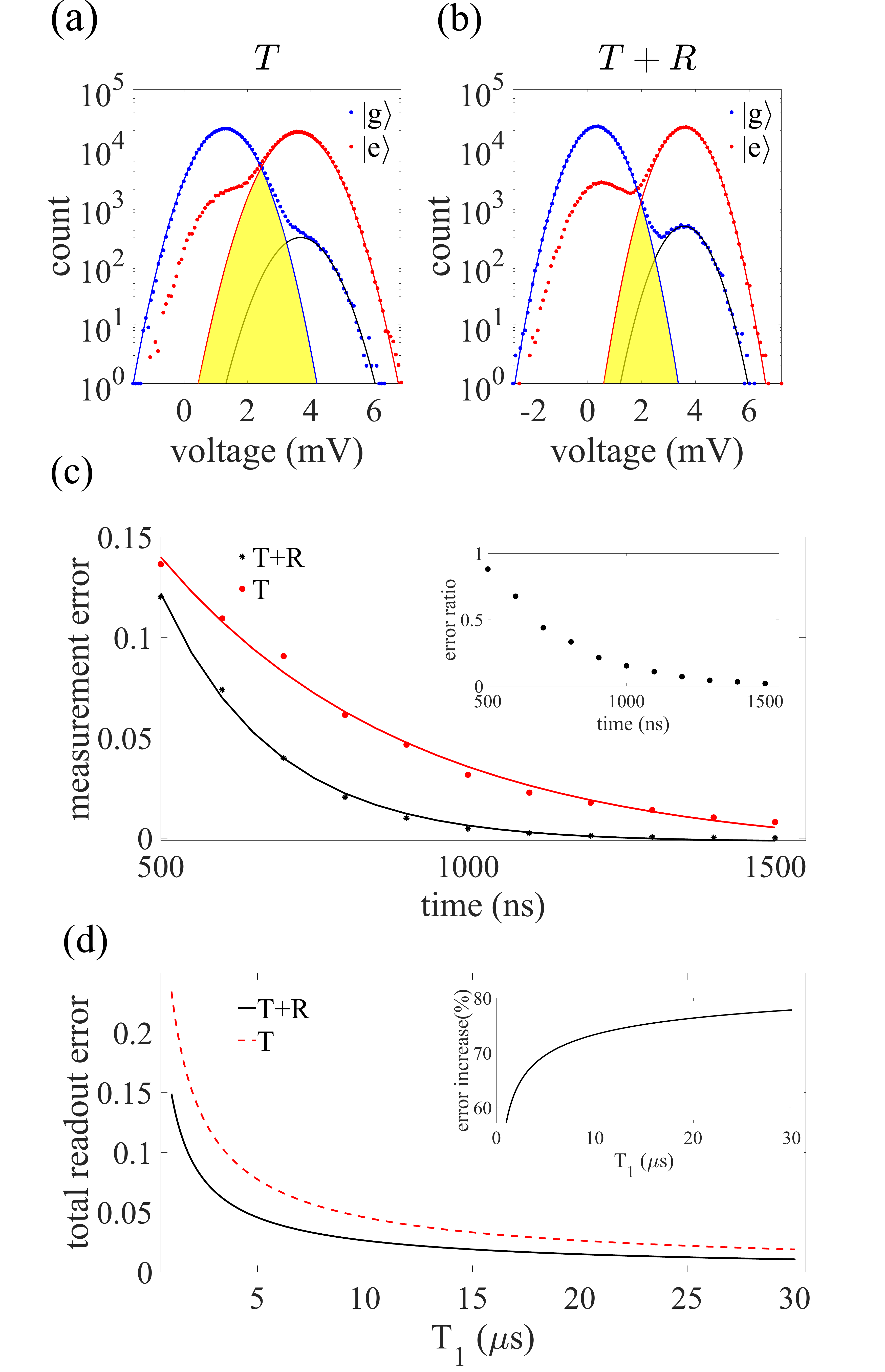}
\caption{Single shot measurement. Histogram plots of the measured cavity response from (a) $T$ path and (b) $T+R$ path, when setting the qubit state to either $\ket{g}$ or $\ket{e}$. The readout time is $900\,$ns. For a given qubit state, the histogram plot normally contains two Gaussian components due to the thermal population and qubit relaxation during the readout process. The measurement error is defined as the overlap of the two Gaussian distributions corresponding to the two pointer states, which is marked with yellow shadow. (c) The measurement errors as a function of readout time for $T$ and $T+R$. The scattered plots are experimental results and the solid lines are from theoretical calculation. Inset of (c) gives the ratio between the measurement errors from $T+R$ and that from $T$. (d) The optimized total error as a function of the qubit relaxation time $T_1$ for $T$ and $T+R$. Inset of (d) gives the error increase when using $T$ for qubit readout, comparing with $T+R$ for varied $T_1$.}
\label{fig_ss}
\end{figure}

In order to evaluate the performance of qubit readout with cavity interference, we implement single shot measurement for the qubit with $\theta_{RT} = 0.11$ from both $T$ and $T+R$. We set the qubit to either $\ket{g}$ or $\ket{e}$,  then record the cavity response to a square pulse with a measurement time $t_m$ in a single shot. Fig.~\ref{fig_ss}(a) and (b) are histogram plots of the single shot measurement from $T$ and $T+R$, respectively. When setting the qubit state to either $\ket{g}$ or $\ket{e}$, the measured voltage distribution contains two Gaussian components, corresponding to a projected state of $\ket{g}$ or $\ket{e}$ by the measurement pulse. The qubit state is assigned to $\ket{g}$ or $\ket{e}$ by comparing the measured cavity response with a threshold voltage $V_{th}$. One could find that there is considerable error probability of assigning the qubit to an incorrect state. The readout error is defined as $\varepsilon = P(e|g)+P(g|e)$, where $P(i|j)$ refers to the probability of assigning the qubit to $\ket{i}$ when the qubit is prepared to $\ket{j}$. Several reasons could contribute to the readout error, including the overlap of the measured cavity response, the qubit relaxation and the non-zero thermal population \cite{Delsing16,Mottonen19}.
The thermal population and qubit relaxation induced readout errors can be suppressed by improving the thermal anchoring and the qubit quality \cite{Oliver2019}, which is beyond the interest of this work. Here we focus on the measurement error, which is induced by the non-zero overlap of the cavity responses and is critically related to the measurement circuit and parameters.

In the experiment we define the measurement error as the overlap of the two Gaussian distributions corresponding to the qubit state which is prepared and measured both in $\ket{g}$ or both in $\ket{e}$. as illustrated in Fig.~\ref{fig_ss}(a) and (b), from which we have the measurement error of $4.8\%$ for $T$, but only $1\%$ for $T+R$. It clearly shows the improvement on the fidelity of single shot qubit readout with path interference. To explicitly demonstrates the suppression of measurement error with path interference, we further compute the measurement error with varied measurement time for both $T$ and $T+R$, which is shown in Fig.~\ref{fig_ss}(c) as scattered plots. Benefiting from the enlarged distance and thus smaller overlap between the two pointer states, the measurement error from $T+R$ is always smaller than that from $T$. The ratio between them is shown in inset of Fig.~\ref{fig_ss}(c). As the measurement time increases, the measurement error from $T+R$ reduces much faster than that from $T$.

We further investigate the effect of path interference on the total readout error. 
Intuitively, if the distance of the cavity responses increases by a factor of $\beta$, the measure time can be reduced by a factor of $1/\beta^2$ to yield the same measurement error. Therefore the total readout error could be reduced considering the reduced qubit relaxation induced error. In fact, for a given measurement condition, there exists an optimized readout time, which shall be long enough to keep a small measurement error, while not leading to a significant qubit relaxation error, and thus a minimum total readout error. Fig.~\ref{fig_ss}(d) gives such an optimized readout error with varied $\ket{e}$ state relaxation time $T_1$, with a photon number in the readout cavity of about $20$. It can be seen that measuring the qubit state with path interference $T+R$ yields a considerably smaller readout error than that with transmission $T$. For a realistically achievable qubit relaxation time $T_1\sim 30 \mu s$ \cite{Oliver2020,Houck2020}, one can reach a readout fidelity better than $99\%$ when measuring with $T+R$, whereas measuring with transmission $T$ bears up to $80\%$ larger readout error.

\section{\label{main:conclusion}Conclusion}

As a summary, we have introduced a new scheme to improve the superconducting qubit readout fidelity with path interference. Our scheme is based on the constructive interference of transmission and reflection of the readout cavity. The cavity photons can be effectively extracted, which leads to a larger separation between the pointer states corresponding to the qubit state $\ket{g}$ and $\ket{e}$. We have developed a method to precisely measure the relative phase between cavity transmission and reflection, which is a critical parameter for the interference process. For the single shot qubit state measurement, we have observed a significant suppression on both measurement error and total readout error when using cavity interference as the readout signal instead of the output from $T$. 
In order to realize simultaneous improvement for multiple cavities coupled to a common transmission line, one could take the distance between adjacent cavities along the transmission line as $\lambda/2$, where $\lambda$ corresponds to the mean value of the resonant frequencies of the cavities, and use a cryogenic phase shifter to realize constructive interference~\cite{Bradley2018,Mikko2017,Mikko2020,Naaman2017}.
As a general scheme to effectively extract the cavity photons, our method can be combined with other readout optimization methods to further improve the performance of superconducting qubit state readout.

This work was supported by the Beijing Academy of quantum information science, the Frontier Science Center for Quantum Information of the Ministry of Education of China through the Tsinghua University Initiative Scientific Research Program, the National Natural Science Foundation of China under Grant No.11874235, the National key Research and Development Program of China (2016YFA0301902, 2020YFA0309500),  Y.K.W. acknowledges support from Shuimu Tsinghua Scholar Program and the International Postdoctoral Exchange Fellowship Program.


\bibliography{ref}

\begin{thebibliography}{42}%
\makeatletter
\providecommand \@ifxundefined [1]{%
 \@ifx{#1\undefined}
}%
\providecommand \@ifnum [1]{%
 \ifnum #1\expandafter \@firstoftwo
 \else \expandafter \@secondoftwo
 \fi
}%
\providecommand \@ifx [1]{%
 \ifx #1\expandafter \@firstoftwo
 \else \expandafter \@secondoftwo
 \fi
}%
\providecommand \natexlab [1]{#1}%
\providecommand \enquote  [1]{``#1''}%
\providecommand \bibnamefont  [1]{#1}%
\providecommand \bibfnamefont [1]{#1}%
\providecommand \citenamefont [1]{#1}%
\providecommand \href@noop [0]{\@secondoftwo}%
\providecommand \href [0]{\begingroup \@sanitize@url \@href}%
\providecommand \@href[1]{\@@startlink{#1}\@@href}%
\providecommand \@@href[1]{\endgroup#1\@@endlink}%
\providecommand \@sanitize@url [0]{\catcode `\\12\catcode `\$12\catcode
  `\&12\catcode `\#12\catcode `\^12\catcode `\_12\catcode `\%12\relax}%
\providecommand \@@startlink[1]{}%
\providecommand \@@endlink[0]{}%
\providecommand \url  [0]{\begingroup\@sanitize@url \@url }%
\providecommand \@url [1]{\endgroup\@href {#1}{\urlprefix }}%
\providecommand \urlprefix  [0]{URL }%
\providecommand \Eprint [0]{\href }%
\providecommand \doibase [0]{https://doi.org/}%
\providecommand \selectlanguage [0]{\@gobble}%
\providecommand \bibinfo  [0]{\@secondoftwo}%
\providecommand \bibfield  [0]{\@secondoftwo}%
\providecommand \translation [1]{[#1]}%
\providecommand \BibitemOpen [0]{}%
\providecommand \bibitemStop [0]{}%
\providecommand \bibitemNoStop [0]{.\EOS\space}%
\providecommand \EOS [0]{\spacefactor3000\relax}%
\providecommand \BibitemShut  [1]{\csname bibitem#1\endcsname}%
\let\auto@bib@innerbib\@empty
\bibitem [{\citenamefont {DiVincenzo}(2000)}]{DiVincenzo2000}%
  \BibitemOpen
  \bibfield  {author} {\bibinfo {author} {\bibfnamefont {D.~P.}\ \bibnamefont
  {DiVincenzo}},\ }\bibfield  {title} {\bibinfo {title} {The physical
  implementation of quantum computation},\ }\href@noop {} {\bibfield  {journal}
  {\bibinfo  {journal} {Fortschritte der Physik}\ }\textbf {\bibinfo {volume}
  {48}},\ \bibinfo {pages} {771} (\bibinfo {year} {2000})}\BibitemShut
  {NoStop}%
\bibitem [{\citenamefont {DiVincenzo}(2009)}]{DiVincenzo2009}%
  \BibitemOpen
  \bibfield  {author} {\bibinfo {author} {\bibfnamefont {D.~P.}\ \bibnamefont
  {DiVincenzo}},\ }\bibfield  {title} {\bibinfo {title} {Fault-tolerant
  architectures for superconducting qubits},\ }\href@noop {} {\bibfield
  {journal} {\bibinfo  {journal} {Physica Scripta}\ }\textbf {\bibinfo {volume}
  {T137}},\ \bibinfo {pages} {014020} (\bibinfo {year} {2009})}\BibitemShut
  {NoStop}%
\bibitem [{\citenamefont {Steffen}\ \emph {et~al.}(2013)\citenamefont
  {Steffen}, \citenamefont {Salathe}, \citenamefont {Oppliger}, \citenamefont
  {Kurpiers}, \citenamefont {Baur}, \citenamefont {Lang}, \citenamefont
  {Eichler}, \citenamefont {Puebla-Hellmann}, \citenamefont {Fedorov},\ and\
  \citenamefont {Wallraff}}]{Wallraff13}%
  \BibitemOpen
  \bibfield  {author} {\bibinfo {author} {\bibfnamefont {L.}~\bibnamefont
  {Steffen}}, \bibinfo {author} {\bibfnamefont {Y.}~\bibnamefont {Salathe}},
  \bibinfo {author} {\bibfnamefont {M.}~\bibnamefont {Oppliger}}, \bibinfo
  {author} {\bibfnamefont {P.}~\bibnamefont {Kurpiers}}, \bibinfo {author}
  {\bibfnamefont {M.}~\bibnamefont {Baur}}, \bibinfo {author} {\bibfnamefont
  {C.}~\bibnamefont {Lang}}, \bibinfo {author} {\bibfnamefont {C.}~\bibnamefont
  {Eichler}}, \bibinfo {author} {\bibfnamefont {G.}~\bibnamefont
  {Puebla-Hellmann}}, \bibinfo {author} {\bibfnamefont {A.}~\bibnamefont
  {Fedorov}},\ and\ \bibinfo {author} {\bibfnamefont {A.}~\bibnamefont
  {Wallraff}},\ }\bibfield  {title} {\bibinfo {title} {Deterministic quantum
  teleportation with feed-forward in a solid state system},\ }\href
  {https://doi.org/10.1038/nature12422} {\bibfield  {journal} {\bibinfo
  {journal} {Nature}\ }\textbf {\bibinfo {volume} {500}},\ \bibinfo {pages}
  {319} (\bibinfo {year} {2013})}\BibitemShut {NoStop}%
\bibitem [{\citenamefont {Barends}\ \emph {et~al.}(2014)\citenamefont
  {Barends}, \citenamefont {Kelly}, \citenamefont {Megrant}, \citenamefont
  {Veitia}, \citenamefont {Sank}, \citenamefont {Jeffrey}, \citenamefont
  {White}, \citenamefont {Mutus}, \citenamefont {Fowler}, \citenamefont
  {Campbell}, \citenamefont {Chen}, \citenamefont {Chen}, \citenamefont
  {Chiaro}, \citenamefont {Dunsworth}, \citenamefont {Neill}, \citenamefont
  {O’Malley}, \citenamefont {Roushan}, \citenamefont {Vainsencher},
  \citenamefont {Wenner}, \citenamefont {Korotkov}, \citenamefont {Cleland},\
  and\ \citenamefont {Martinis}}]{Martinis2014}%
  \BibitemOpen
  \bibfield  {author} {\bibinfo {author} {\bibfnamefont {R.}~\bibnamefont
  {Barends}}, \bibinfo {author} {\bibfnamefont {J.}~\bibnamefont {Kelly}},
  \bibinfo {author} {\bibfnamefont {A.}~\bibnamefont {Megrant}}, \bibinfo
  {author} {\bibfnamefont {A.}~\bibnamefont {Veitia}}, \bibinfo {author}
  {\bibfnamefont {D.}~\bibnamefont {Sank}}, \bibinfo {author} {\bibfnamefont
  {E.}~\bibnamefont {Jeffrey}}, \bibinfo {author} {\bibfnamefont {T.~C.}\
  \bibnamefont {White}}, \bibinfo {author} {\bibfnamefont {J.}~\bibnamefont
  {Mutus}}, \bibinfo {author} {\bibfnamefont {A.~G.}\ \bibnamefont {Fowler}},
  \bibinfo {author} {\bibfnamefont {B.}~\bibnamefont {Campbell}}, \bibinfo
  {author} {\bibfnamefont {Y.}~\bibnamefont {Chen}}, \bibinfo {author}
  {\bibfnamefont {Z.}~\bibnamefont {Chen}}, \bibinfo {author} {\bibfnamefont
  {B.}~\bibnamefont {Chiaro}}, \bibinfo {author} {\bibfnamefont
  {A.}~\bibnamefont {Dunsworth}}, \bibinfo {author} {\bibfnamefont
  {C.}~\bibnamefont {Neill}}, \bibinfo {author} {\bibfnamefont
  {P.}~\bibnamefont {O’Malley}}, \bibinfo {author} {\bibfnamefont
  {P.}~\bibnamefont {Roushan}}, \bibinfo {author} {\bibfnamefont
  {A.}~\bibnamefont {Vainsencher}}, \bibinfo {author} {\bibfnamefont
  {J.}~\bibnamefont {Wenner}}, \bibinfo {author} {\bibfnamefont {A.~N.}\
  \bibnamefont {Korotkov}}, \bibinfo {author} {\bibfnamefont {A.~N.}\
  \bibnamefont {Cleland}},\ and\ \bibinfo {author} {\bibfnamefont {J.~M.}\
  \bibnamefont {Martinis}},\ }\bibfield  {title} {\bibinfo {title}
  {Superconducting quantum circuits at the surface code threshold for fault
  tolerance},\ }\href {https://doi.org/10.1038/nature13171} {\bibfield
  {journal} {\bibinfo  {journal} {Nature}\ }\textbf {\bibinfo {volume} {508}},\
  \bibinfo {pages} {500} (\bibinfo {year} {2014})}\BibitemShut {NoStop}%
\bibitem [{\citenamefont {Kelly}\ \emph {et~al.}(2015)\citenamefont {Kelly},
  \citenamefont {Barends}, \citenamefont {Fowler}, \citenamefont {Megrant},
  \citenamefont {Jeffrey}, \citenamefont {White}, \citenamefont {Sank},
  \citenamefont {Mutus}, \citenamefont {Campbell}, \citenamefont {Chen},
  \citenamefont {Chen}, \citenamefont {Chiaro}, \citenamefont {Dunsworth},
  \citenamefont {Hoi}, \citenamefont {Neill}, \citenamefont {O’Malley},
  \citenamefont {Quintana}, \citenamefont {Roushan}, \citenamefont
  {Vainsencher}, \citenamefont {Wenner}, \citenamefont {Cleland},\ and\
  \citenamefont {Martinis}}]{Martinis15nature}%
  \BibitemOpen
  \bibfield  {author} {\bibinfo {author} {\bibfnamefont {J.}~\bibnamefont
  {Kelly}}, \bibinfo {author} {\bibfnamefont {R.}~\bibnamefont {Barends}},
  \bibinfo {author} {\bibfnamefont {A.~G.}\ \bibnamefont {Fowler}}, \bibinfo
  {author} {\bibfnamefont {A.}~\bibnamefont {Megrant}}, \bibinfo {author}
  {\bibfnamefont {E.}~\bibnamefont {Jeffrey}}, \bibinfo {author} {\bibfnamefont
  {T.~C.}\ \bibnamefont {White}}, \bibinfo {author} {\bibfnamefont
  {D.}~\bibnamefont {Sank}}, \bibinfo {author} {\bibfnamefont {J.~Y.}\
  \bibnamefont {Mutus}}, \bibinfo {author} {\bibfnamefont {B.}~\bibnamefont
  {Campbell}}, \bibinfo {author} {\bibfnamefont {Y.}~\bibnamefont {Chen}},
  \bibinfo {author} {\bibfnamefont {Z.}~\bibnamefont {Chen}}, \bibinfo {author}
  {\bibfnamefont {B.}~\bibnamefont {Chiaro}}, \bibinfo {author} {\bibfnamefont
  {A.}~\bibnamefont {Dunsworth}}, \bibinfo {author} {\bibfnamefont {I.~C.}\
  \bibnamefont {Hoi}}, \bibinfo {author} {\bibfnamefont {C.}~\bibnamefont
  {Neill}}, \bibinfo {author} {\bibfnamefont {P.~J.~J.}\ \bibnamefont
  {O’Malley}}, \bibinfo {author} {\bibfnamefont {C.}~\bibnamefont
  {Quintana}}, \bibinfo {author} {\bibfnamefont {P.}~\bibnamefont {Roushan}},
  \bibinfo {author} {\bibfnamefont {A.}~\bibnamefont {Vainsencher}}, \bibinfo
  {author} {\bibfnamefont {J.}~\bibnamefont {Wenner}}, \bibinfo {author}
  {\bibfnamefont {A.~N.}\ \bibnamefont {Cleland}},\ and\ \bibinfo {author}
  {\bibfnamefont {J.~M.}\ \bibnamefont {Martinis}},\ }\bibfield  {title}
  {\bibinfo {title} {State preservation by repetitive error detection in a
  superconducting quantum circuit},\ }\href
  {https://doi.org/10.1038/nature14270} {\bibfield  {journal} {\bibinfo
  {journal} {Nature}\ }\textbf {\bibinfo {volume} {519}},\ \bibinfo {pages}
  {66} (\bibinfo {year} {2015})}\BibitemShut {NoStop}%
\bibitem [{\citenamefont {Johnson}\ \emph {et~al.}(2012)\citenamefont
  {Johnson}, \citenamefont {Macklin}, \citenamefont {Slichter}, \citenamefont
  {Vijay}, \citenamefont {Weingarten}, \citenamefont {Clarke},\ and\
  \citenamefont {Siddiqi}}]{Siddiqi2012}%
  \BibitemOpen
  \bibfield  {author} {\bibinfo {author} {\bibfnamefont {J.~E.}\ \bibnamefont
  {Johnson}}, \bibinfo {author} {\bibfnamefont {C.}~\bibnamefont {Macklin}},
  \bibinfo {author} {\bibfnamefont {D.~H.}\ \bibnamefont {Slichter}}, \bibinfo
  {author} {\bibfnamefont {R.}~\bibnamefont {Vijay}}, \bibinfo {author}
  {\bibfnamefont {E.~B.}\ \bibnamefont {Weingarten}}, \bibinfo {author}
  {\bibfnamefont {J.}~\bibnamefont {Clarke}},\ and\ \bibinfo {author}
  {\bibfnamefont {I.}~\bibnamefont {Siddiqi}},\ }\bibfield  {title} {\bibinfo
  {title} {Heralded state preparation in a superconducting qubit},\ }\href@noop
  {} {\bibfield  {journal} {\bibinfo  {journal} {Phys. Rev. Lett.}\ }\textbf
  {\bibinfo {volume} {109}},\ \bibinfo {pages} {050506} (\bibinfo {year}
  {2012})}\BibitemShut {NoStop}%
\bibitem [{\citenamefont {Rist\`e}\ \emph {et~al.}(2012)\citenamefont
  {Rist\`e}, \citenamefont {van Leeuwen}, \citenamefont {Ku}, \citenamefont
  {Lehnert},\ and\ \citenamefont {DiCarlo}}]{DiCarlo2012}%
  \BibitemOpen
  \bibfield  {author} {\bibinfo {author} {\bibfnamefont {D.}~\bibnamefont
  {Rist\`e}}, \bibinfo {author} {\bibfnamefont {J.~G.}\ \bibnamefont {van
  Leeuwen}}, \bibinfo {author} {\bibfnamefont {H.-S.}\ \bibnamefont {Ku}},
  \bibinfo {author} {\bibfnamefont {K.~W.}\ \bibnamefont {Lehnert}},\ and\
  \bibinfo {author} {\bibfnamefont {L.}~\bibnamefont {DiCarlo}},\ }\bibfield
  {title} {\bibinfo {title} {Initialization by measurement of a superconducting
  quantum bit circuit},\ }\href@noop {} {\bibfield  {journal} {\bibinfo
  {journal} {Phys. Rev. Lett.}\ }\textbf {\bibinfo {volume} {109}},\ \bibinfo
  {pages} {050507} (\bibinfo {year} {2012})}\BibitemShut {NoStop}%
\bibitem [{\citenamefont {Geerlings}\ \emph {et~al.}(2013)\citenamefont
  {Geerlings}, \citenamefont {Leghtas}, \citenamefont {Pop}, \citenamefont
  {Shankar}, \citenamefont {Frunzio}, \citenamefont {Schoelkopf}, \citenamefont
  {Mirrahimi},\ and\ \citenamefont {Devoret}}]{Devoret2013}%
  \BibitemOpen
  \bibfield  {author} {\bibinfo {author} {\bibfnamefont {K.}~\bibnamefont
  {Geerlings}}, \bibinfo {author} {\bibfnamefont {Z.}~\bibnamefont {Leghtas}},
  \bibinfo {author} {\bibfnamefont {I.~M.}\ \bibnamefont {Pop}}, \bibinfo
  {author} {\bibfnamefont {S.}~\bibnamefont {Shankar}}, \bibinfo {author}
  {\bibfnamefont {L.}~\bibnamefont {Frunzio}}, \bibinfo {author} {\bibfnamefont
  {R.~J.}\ \bibnamefont {Schoelkopf}}, \bibinfo {author} {\bibfnamefont
  {M.}~\bibnamefont {Mirrahimi}},\ and\ \bibinfo {author} {\bibfnamefont
  {M.~H.}\ \bibnamefont {Devoret}},\ }\bibfield  {title} {\bibinfo {title}
  {Demonstrating a driven reset protocol for a superconducting qubit},\
  }\href@noop {} {\bibfield  {journal} {\bibinfo  {journal} {Phys. Rev. Lett.}\
  }\textbf {\bibinfo {volume} {110}},\ \bibinfo {pages} {120501} (\bibinfo
  {year} {2013})}\BibitemShut {NoStop}%
\bibitem [{\citenamefont {Magnard}\ \emph {et~al.}(2018)\citenamefont
  {Magnard}, \citenamefont {Kurpiers}, \citenamefont {Royer}, \citenamefont
  {Walter}, \citenamefont {Besse}, \citenamefont {Gasparinetti}, \citenamefont
  {Pechal}, \citenamefont {Heinsoo}, \citenamefont {Storz}, \citenamefont
  {Blais},\ and\ \citenamefont {Wallraff}}]{Wallraff2018}%
  \BibitemOpen
  \bibfield  {author} {\bibinfo {author} {\bibfnamefont {P.}~\bibnamefont
  {Magnard}}, \bibinfo {author} {\bibfnamefont {P.}~\bibnamefont {Kurpiers}},
  \bibinfo {author} {\bibfnamefont {B.}~\bibnamefont {Royer}}, \bibinfo
  {author} {\bibfnamefont {T.}~\bibnamefont {Walter}}, \bibinfo {author}
  {\bibfnamefont {J.-C.}\ \bibnamefont {Besse}}, \bibinfo {author}
  {\bibfnamefont {S.}~\bibnamefont {Gasparinetti}}, \bibinfo {author}
  {\bibfnamefont {M.}~\bibnamefont {Pechal}}, \bibinfo {author} {\bibfnamefont
  {J.}~\bibnamefont {Heinsoo}}, \bibinfo {author} {\bibfnamefont
  {S.}~\bibnamefont {Storz}}, \bibinfo {author} {\bibfnamefont
  {A.}~\bibnamefont {Blais}},\ and\ \bibinfo {author} {\bibfnamefont
  {A.}~\bibnamefont {Wallraff}},\ }\bibfield  {title} {\bibinfo {title} {Fast
  and unconditional all-microwave reset of a superconducting qubit},\
  }\href@noop {} {\bibfield  {journal} {\bibinfo  {journal} {Phys. Rev. Lett.}\
  }\textbf {\bibinfo {volume} {121}},\ \bibinfo {pages} {060502} (\bibinfo
  {year} {2018})}\BibitemShut {NoStop}%
\bibitem [{\citenamefont {Arute}\ \emph {et~al.}(2019)\citenamefont {Arute},
  \citenamefont {Arya}, \citenamefont {Babbush}, \citenamefont {Bacon},
  \citenamefont {Bardin}, \citenamefont {Barends}, \citenamefont {Biswas},
  \citenamefont {Boixo}, \citenamefont {Brandao}, \citenamefont {Buell},
  \citenamefont {Burkett}, \citenamefont {Chen}, \citenamefont {Chen},
  \citenamefont {Chiaro}, \citenamefont {Collins}, \citenamefont {Courtney},
  \citenamefont {Dunsworth}, \citenamefont {Farhi}, \citenamefont {Foxen},
  \citenamefont {Fowler}, \citenamefont {Gidney}, \citenamefont {Giustina},
  \citenamefont {Graff}, \citenamefont {Guerin}, \citenamefont {Habegger},
  \citenamefont {Harrigan}, \citenamefont {Hartmann}, \citenamefont {Ho},
  \citenamefont {Hoffmann}, \citenamefont {Huang}, \citenamefont {Humble},
  \citenamefont {Isakov}, \citenamefont {Jeffrey}, \citenamefont {Jiang},
  \citenamefont {Kafri}, \citenamefont {Kechedzhi}, \citenamefont {Kelly},
  \citenamefont {Klimov}, \citenamefont {Knysh}, \citenamefont {Korotkov},
  \citenamefont {Kostritsa}, \citenamefont {Landhuis}, \citenamefont
  {Lindmark}, \citenamefont {Lucero}, \citenamefont {Lyakh}, \citenamefont
  {Mandrà}, \citenamefont {McClean}, \citenamefont {McEwen}, \citenamefont
  {Megrant}, \citenamefont {Mi}, \citenamefont {Michielsen}, \citenamefont
  {Mohseni}, \citenamefont {Mutus}, \citenamefont {Naaman}, \citenamefont
  {Neeley}, \citenamefont {Neill}, \citenamefont {Niu}, \citenamefont {Ostby},
  \citenamefont {Petukhov}, \citenamefont {Platt}, \citenamefont {Quintana},
  \citenamefont {Rieffel}, \citenamefont {Roushan}, \citenamefont {Rubin},
  \citenamefont {Sank}, \citenamefont {Satzinger}, \citenamefont {Smelyanskiy},
  \citenamefont {Sung}, \citenamefont {Trevithick}, \citenamefont
  {Vainsencher}, \citenamefont {Villalonga}, \citenamefont {White},
  \citenamefont {Yao}, \citenamefont {Yeh}, \citenamefont {Zalcman},
  \citenamefont {Neven},\ and\ \citenamefont {Martinis}}]{Martinis19nature}%
  \BibitemOpen
  \bibfield  {author} {\bibinfo {author} {\bibfnamefont {F.}~\bibnamefont
  {Arute}}, \bibinfo {author} {\bibfnamefont {K.}~\bibnamefont {Arya}},
  \bibinfo {author} {\bibfnamefont {R.}~\bibnamefont {Babbush}}, \bibinfo
  {author} {\bibfnamefont {D.}~\bibnamefont {Bacon}}, \bibinfo {author}
  {\bibfnamefont {J.~C.}\ \bibnamefont {Bardin}}, \bibinfo {author}
  {\bibfnamefont {R.}~\bibnamefont {Barends}}, \bibinfo {author} {\bibfnamefont
  {R.}~\bibnamefont {Biswas}}, \bibinfo {author} {\bibfnamefont
  {S.}~\bibnamefont {Boixo}}, \bibinfo {author} {\bibfnamefont {F.~G. S.~L.}\
  \bibnamefont {Brandao}}, \bibinfo {author} {\bibfnamefont {D.~A.}\
  \bibnamefont {Buell}}, \bibinfo {author} {\bibfnamefont {B.}~\bibnamefont
  {Burkett}}, \bibinfo {author} {\bibfnamefont {Y.}~\bibnamefont {Chen}},
  \bibinfo {author} {\bibfnamefont {Z.}~\bibnamefont {Chen}}, \bibinfo {author}
  {\bibfnamefont {B.}~\bibnamefont {Chiaro}}, \bibinfo {author} {\bibfnamefont
  {R.}~\bibnamefont {Collins}}, \bibinfo {author} {\bibfnamefont
  {W.}~\bibnamefont {Courtney}}, \bibinfo {author} {\bibfnamefont
  {A.}~\bibnamefont {Dunsworth}}, \bibinfo {author} {\bibfnamefont
  {E.}~\bibnamefont {Farhi}}, \bibinfo {author} {\bibfnamefont
  {B.}~\bibnamefont {Foxen}}, \bibinfo {author} {\bibfnamefont
  {A.}~\bibnamefont {Fowler}}, \bibinfo {author} {\bibfnamefont
  {C.}~\bibnamefont {Gidney}}, \bibinfo {author} {\bibfnamefont
  {M.}~\bibnamefont {Giustina}}, \bibinfo {author} {\bibfnamefont
  {R.}~\bibnamefont {Graff}}, \bibinfo {author} {\bibfnamefont
  {K.}~\bibnamefont {Guerin}}, \bibinfo {author} {\bibfnamefont
  {S.}~\bibnamefont {Habegger}}, \bibinfo {author} {\bibfnamefont {M.~P.}\
  \bibnamefont {Harrigan}}, \bibinfo {author} {\bibfnamefont {M.~J.}\
  \bibnamefont {Hartmann}}, \bibinfo {author} {\bibfnamefont {A.}~\bibnamefont
  {Ho}}, \bibinfo {author} {\bibfnamefont {M.}~\bibnamefont {Hoffmann}},
  \bibinfo {author} {\bibfnamefont {T.}~\bibnamefont {Huang}}, \bibinfo
  {author} {\bibfnamefont {T.~S.}\ \bibnamefont {Humble}}, \bibinfo {author}
  {\bibfnamefont {S.~V.}\ \bibnamefont {Isakov}}, \bibinfo {author}
  {\bibfnamefont {E.}~\bibnamefont {Jeffrey}}, \bibinfo {author} {\bibfnamefont
  {Z.}~\bibnamefont {Jiang}}, \bibinfo {author} {\bibfnamefont
  {D.}~\bibnamefont {Kafri}}, \bibinfo {author} {\bibfnamefont
  {K.}~\bibnamefont {Kechedzhi}}, \bibinfo {author} {\bibfnamefont
  {J.}~\bibnamefont {Kelly}}, \bibinfo {author} {\bibfnamefont {P.~V.}\
  \bibnamefont {Klimov}}, \bibinfo {author} {\bibfnamefont {S.}~\bibnamefont
  {Knysh}}, \bibinfo {author} {\bibfnamefont {A.}~\bibnamefont {Korotkov}},
  \bibinfo {author} {\bibfnamefont {F.}~\bibnamefont {Kostritsa}}, \bibinfo
  {author} {\bibfnamefont {D.}~\bibnamefont {Landhuis}}, \bibinfo {author}
  {\bibfnamefont {M.}~\bibnamefont {Lindmark}}, \bibinfo {author}
  {\bibfnamefont {E.}~\bibnamefont {Lucero}}, \bibinfo {author} {\bibfnamefont
  {D.}~\bibnamefont {Lyakh}}, \bibinfo {author} {\bibfnamefont
  {S.}~\bibnamefont {Mandrà}}, \bibinfo {author} {\bibfnamefont {J.~R.}\
  \bibnamefont {McClean}}, \bibinfo {author} {\bibfnamefont {M.}~\bibnamefont
  {McEwen}}, \bibinfo {author} {\bibfnamefont {A.}~\bibnamefont {Megrant}},
  \bibinfo {author} {\bibfnamefont {X.}~\bibnamefont {Mi}}, \bibinfo {author}
  {\bibfnamefont {K.}~\bibnamefont {Michielsen}}, \bibinfo {author}
  {\bibfnamefont {M.}~\bibnamefont {Mohseni}}, \bibinfo {author} {\bibfnamefont
  {J.}~\bibnamefont {Mutus}}, \bibinfo {author} {\bibfnamefont
  {O.}~\bibnamefont {Naaman}}, \bibinfo {author} {\bibfnamefont
  {M.}~\bibnamefont {Neeley}}, \bibinfo {author} {\bibfnamefont
  {C.}~\bibnamefont {Neill}}, \bibinfo {author} {\bibfnamefont {M.~Y.}\
  \bibnamefont {Niu}}, \bibinfo {author} {\bibfnamefont {E.}~\bibnamefont
  {Ostby}}, \bibinfo {author} {\bibfnamefont {A.}~\bibnamefont {Petukhov}},
  \bibinfo {author} {\bibfnamefont {J.~C.}\ \bibnamefont {Platt}}, \bibinfo
  {author} {\bibfnamefont {C.}~\bibnamefont {Quintana}}, \bibinfo {author}
  {\bibfnamefont {E.~G.}\ \bibnamefont {Rieffel}}, \bibinfo {author}
  {\bibfnamefont {P.}~\bibnamefont {Roushan}}, \bibinfo {author} {\bibfnamefont
  {N.~C.}\ \bibnamefont {Rubin}}, \bibinfo {author} {\bibfnamefont
  {D.}~\bibnamefont {Sank}}, \bibinfo {author} {\bibfnamefont {K.~J.}\
  \bibnamefont {Satzinger}}, \bibinfo {author} {\bibfnamefont {V.}~\bibnamefont
  {Smelyanskiy}}, \bibinfo {author} {\bibfnamefont {K.~J.}\ \bibnamefont
  {Sung}}, \bibinfo {author} {\bibfnamefont {M.~D.}\ \bibnamefont
  {Trevithick}}, \bibinfo {author} {\bibfnamefont {A.}~\bibnamefont
  {Vainsencher}}, \bibinfo {author} {\bibfnamefont {B.}~\bibnamefont
  {Villalonga}}, \bibinfo {author} {\bibfnamefont {T.}~\bibnamefont {White}},
  \bibinfo {author} {\bibfnamefont {Z.~J.}\ \bibnamefont {Yao}}, \bibinfo
  {author} {\bibfnamefont {P.}~\bibnamefont {Yeh}}, \bibinfo {author}
  {\bibfnamefont {A.}~\bibnamefont {Zalcman}}, \bibinfo {author} {\bibfnamefont
  {H.}~\bibnamefont {Neven}},\ and\ \bibinfo {author} {\bibfnamefont {J.~M.}\
  \bibnamefont {Martinis}},\ }\bibfield  {title} {\bibinfo {title} {Quantum
  supremacy using a programmable superconducting processor},\ }\href
  {https://doi.org/10.1038/s41586-019-1666-5} {\bibfield  {journal} {\bibinfo
  {journal} {Nature}\ }\textbf {\bibinfo {volume} {574}},\ \bibinfo {pages}
  {505} (\bibinfo {year} {2019})}\BibitemShut {NoStop}%
\bibitem [{\citenamefont {Kandala}\ \emph {et~al.}(2019)\citenamefont
  {Kandala}, \citenamefont {Temme}, \citenamefont {Córcoles}, \citenamefont
  {Mezzacapo}, \citenamefont {Chow},\ and\ \citenamefont {Gambetta}}]{IBM2019}%
  \BibitemOpen
  \bibfield  {author} {\bibinfo {author} {\bibfnamefont {A.}~\bibnamefont
  {Kandala}}, \bibinfo {author} {\bibfnamefont {K.}~\bibnamefont {Temme}},
  \bibinfo {author} {\bibfnamefont {A.~D.}\ \bibnamefont {Córcoles}}, \bibinfo
  {author} {\bibfnamefont {A.}~\bibnamefont {Mezzacapo}}, \bibinfo {author}
  {\bibfnamefont {J.~M.}\ \bibnamefont {Chow}},\ and\ \bibinfo {author}
  {\bibfnamefont {J.~M.}\ \bibnamefont {Gambetta}},\ }\bibfield  {title}
  {\bibinfo {title} {Error mitigation extends the computational reach of a
  noisy quantum processor},\ }\href {https://doi.org/10.1038/s41586-019-1040-7}
  {\bibfield  {journal} {\bibinfo  {journal} {Nature}\ }\textbf {\bibinfo
  {volume} {567}},\ \bibinfo {pages} {491} (\bibinfo {year}
  {2019})}\BibitemShut {NoStop}%
\bibitem [{\citenamefont {Tannu}\ and\ \citenamefont
  {Qureshi}(2019{\natexlab{a}})}]{Tannu2019}%
  \BibitemOpen
  \bibfield  {author} {\bibinfo {author} {\bibfnamefont {S.~S.}\ \bibnamefont
  {Tannu}}\ and\ \bibinfo {author} {\bibfnamefont {M.~K.}\ \bibnamefont
  {Qureshi}},\ }\bibfield  {title} {\bibinfo {title} {Not all qubits are
  created equal: A case for variability-aware policies for nisq-era quantum
  computers},\ }in\ \href {https://doi.org/10.1145/3297858.3304007} {\emph
  {\bibinfo {booktitle} {Proceedings of the Twenty-Fourth International
  Conference on Architectural Support for Programming Languages and Operating
  Systems}}},\ \bibinfo {series and number} {ASPLOS ’19}\ (\bibinfo
  {publisher} {Association for Computing Machinery},\ \bibinfo {address} {New
  York, NY, USA},\ \bibinfo {year} {2019})\ p.\ \bibinfo {pages}
  {987–999}\BibitemShut {NoStop}%
\bibitem [{\citenamefont {Tannu}\ and\ \citenamefont
  {Qureshi}(2019{\natexlab{b}})}]{Tannu2019a}%
  \BibitemOpen
  \bibfield  {author} {\bibinfo {author} {\bibfnamefont {S.~S.}\ \bibnamefont
  {Tannu}}\ and\ \bibinfo {author} {\bibfnamefont {M.~K.}\ \bibnamefont
  {Qureshi}},\ }\bibfield  {title} {\bibinfo {title} {Mitigating measurement
  errors in quantum computers by exploiting state-dependent bias},\ }in\
  \href@noop {} {\emph {\bibinfo {booktitle} {Proceedings of the 52nd Annual
  IEEE/ACM International Symposium on Microarchitecture}}},\ \bibinfo {series
  and number} {MICRO '52}\ (\bibinfo  {publisher} {Association for Computing
  Machinery},\ \bibinfo {address} {New York, NY, USA},\ \bibinfo {year}
  {2019})\ p.\ \bibinfo {pages} {279–290}\BibitemShut {NoStop}%
\bibitem [{\citenamefont {Lupaşcu}\ \emph {et~al.}(2007)\citenamefont
  {Lupaşcu}, \citenamefont {Saito}, \citenamefont {Picot}, \citenamefont
  {de~Groot}, \citenamefont {Harmans},\ and\ \citenamefont
  {Mooij}}]{Mooij2007}%
  \BibitemOpen
  \bibfield  {author} {\bibinfo {author} {\bibfnamefont {A.}~\bibnamefont
  {Lupaşcu}}, \bibinfo {author} {\bibfnamefont {S.}~\bibnamefont {Saito}},
  \bibinfo {author} {\bibfnamefont {T.}~\bibnamefont {Picot}}, \bibinfo
  {author} {\bibfnamefont {P.~C.}\ \bibnamefont {de~Groot}}, \bibinfo {author}
  {\bibfnamefont {C.~J. P.~M.}\ \bibnamefont {Harmans}},\ and\ \bibinfo
  {author} {\bibfnamefont {J.~E.}\ \bibnamefont {Mooij}},\ }\bibfield  {title}
  {\bibinfo {title} {Quantum non-demolition measurement of a superconducting
  two-level system},\ }\href {https://doi.org/10.1038/nphys509} {\bibfield
  {journal} {\bibinfo  {journal} {Nature Physics}\ }\textbf {\bibinfo {volume}
  {3}},\ \bibinfo {pages} {119} (\bibinfo {year} {2007})}\BibitemShut {NoStop}%
\bibitem [{\citenamefont {Picot}\ \emph {et~al.}(2010)\citenamefont {Picot},
  \citenamefont {Schouten}, \citenamefont {Harmans},\ and\ \citenamefont
  {Mooij}}]{Mooij2010}%
  \BibitemOpen
  \bibfield  {author} {\bibinfo {author} {\bibfnamefont {T.}~\bibnamefont
  {Picot}}, \bibinfo {author} {\bibfnamefont {R.}~\bibnamefont {Schouten}},
  \bibinfo {author} {\bibfnamefont {C.~J. P.~M.}\ \bibnamefont {Harmans}},\
  and\ \bibinfo {author} {\bibfnamefont {J.~E.}\ \bibnamefont {Mooij}},\
  }\bibfield  {title} {\bibinfo {title} {Quantum nondemolition measurement of a
  superconducting qubit in the weakly projective regime},\ }\href
  {https://doi.org/10.1103/PhysRevLett.105.040506} {\bibfield  {journal}
  {\bibinfo  {journal} {Physical Review Letters}\ }\textbf {\bibinfo {volume}
  {105}},\ \bibinfo {pages} {040506} (\bibinfo {year} {2010})}\BibitemShut
  {NoStop}%
\bibitem [{\citenamefont {Nakajima}\ \emph {et~al.}(2019)\citenamefont
  {Nakajima}, \citenamefont {Noiri}, \citenamefont {Yoneda}, \citenamefont
  {Delbecq}, \citenamefont {Stano}, \citenamefont {Otsuka}, \citenamefont
  {Takeda}, \citenamefont {Amaha}, \citenamefont {Allison}, \citenamefont
  {Kawasaki}, \citenamefont {Ludwig}, \citenamefont {Wieck}, \citenamefont
  {Loss},\ and\ \citenamefont {Tarucha}}]{Tarucha2019}%
  \BibitemOpen
  \bibfield  {author} {\bibinfo {author} {\bibfnamefont {T.}~\bibnamefont
  {Nakajima}}, \bibinfo {author} {\bibfnamefont {A.}~\bibnamefont {Noiri}},
  \bibinfo {author} {\bibfnamefont {J.}~\bibnamefont {Yoneda}}, \bibinfo
  {author} {\bibfnamefont {M.~R.}\ \bibnamefont {Delbecq}}, \bibinfo {author}
  {\bibfnamefont {P.}~\bibnamefont {Stano}}, \bibinfo {author} {\bibfnamefont
  {T.}~\bibnamefont {Otsuka}}, \bibinfo {author} {\bibfnamefont
  {K.}~\bibnamefont {Takeda}}, \bibinfo {author} {\bibfnamefont
  {S.}~\bibnamefont {Amaha}}, \bibinfo {author} {\bibfnamefont
  {G.}~\bibnamefont {Allison}}, \bibinfo {author} {\bibfnamefont
  {K.}~\bibnamefont {Kawasaki}}, \bibinfo {author} {\bibfnamefont
  {A.}~\bibnamefont {Ludwig}}, \bibinfo {author} {\bibfnamefont {A.~D.}\
  \bibnamefont {Wieck}}, \bibinfo {author} {\bibfnamefont {D.}~\bibnamefont
  {Loss}},\ and\ \bibinfo {author} {\bibfnamefont {S.}~\bibnamefont
  {Tarucha}},\ }\bibfield  {title} {\bibinfo {title} {Quantum non-demolition
  measurement of an electron spin qubit},\ }\href
  {https://doi.org/10.1038/s41565-019-0426-x} {\bibfield  {journal} {\bibinfo
  {journal} {Nature Nanotechnology}\ }\textbf {\bibinfo {volume} {14}},\
  \bibinfo {pages} {555} (\bibinfo {year} {2019})}\BibitemShut {NoStop}%
\bibitem [{\citenamefont {Raha}\ \emph {et~al.}(2020)\citenamefont {Raha},
  \citenamefont {Chen}, \citenamefont {Phenicie}, \citenamefont {Ourari},
  \citenamefont {Dibos},\ and\ \citenamefont {Thompson}}]{Raha2020}%
  \BibitemOpen
  \bibfield  {author} {\bibinfo {author} {\bibfnamefont {M.}~\bibnamefont
  {Raha}}, \bibinfo {author} {\bibfnamefont {S.}~\bibnamefont {Chen}}, \bibinfo
  {author} {\bibfnamefont {C.~M.}\ \bibnamefont {Phenicie}}, \bibinfo {author}
  {\bibfnamefont {S.}~\bibnamefont {Ourari}}, \bibinfo {author} {\bibfnamefont
  {A.~M.}\ \bibnamefont {Dibos}},\ and\ \bibinfo {author} {\bibfnamefont
  {J.~D.}\ \bibnamefont {Thompson}},\ }\bibfield  {title} {\bibinfo {title}
  {Optical quantum nondemolition measurement of a single rare earth ion
  qubit},\ }\href {https://doi.org/10.1038/s41467-020-15138-7} {\bibfield
  {journal} {\bibinfo  {journal} {Nature Communications}\ }\textbf {\bibinfo
  {volume} {11}},\ \bibinfo {pages} {1605} (\bibinfo {year}
  {2020})}\BibitemShut {NoStop}%
\bibitem [{\citenamefont {Ristè}\ \emph {et~al.}(2015)\citenamefont {Ristè},
  \citenamefont {Poletto}, \citenamefont {Huang}, \citenamefont {Bruno},
  \citenamefont {Vesterinen}, \citenamefont {Saira},\ and\ \citenamefont
  {DiCarlo}}]{DiCarlo15}%
  \BibitemOpen
  \bibfield  {author} {\bibinfo {author} {\bibfnamefont {D.}~\bibnamefont
  {Ristè}}, \bibinfo {author} {\bibfnamefont {S.}~\bibnamefont {Poletto}},
  \bibinfo {author} {\bibfnamefont {M.~Z.}\ \bibnamefont {Huang}}, \bibinfo
  {author} {\bibfnamefont {A.}~\bibnamefont {Bruno}}, \bibinfo {author}
  {\bibfnamefont {V.}~\bibnamefont {Vesterinen}}, \bibinfo {author}
  {\bibfnamefont {O.~P.}\ \bibnamefont {Saira}},\ and\ \bibinfo {author}
  {\bibfnamefont {L.}~\bibnamefont {DiCarlo}},\ }\bibfield  {title} {\bibinfo
  {title} {Detecting bit-flip errors in a logical qubit using stabilizer
  measurements},\ }\href {https://doi.org/10.1038/ncomms7983} {\bibfield
  {journal} {\bibinfo  {journal} {Nature Communications}\ }\textbf {\bibinfo
  {volume} {6}},\ \bibinfo {pages} {6983} (\bibinfo {year} {2015})}\BibitemShut
  {NoStop}%
\bibitem [{\citenamefont {Hacohen-Gourgy}\ \emph {et~al.}(2016)\citenamefont
  {Hacohen-Gourgy}, \citenamefont {Martin}, \citenamefont {Flurin},
  \citenamefont {Ramasesh}, \citenamefont {Whaley},\ and\ \citenamefont
  {Siddiqi}}]{Siddiqi16}%
  \BibitemOpen
  \bibfield  {author} {\bibinfo {author} {\bibfnamefont {S.}~\bibnamefont
  {Hacohen-Gourgy}}, \bibinfo {author} {\bibfnamefont {L.~S.}\ \bibnamefont
  {Martin}}, \bibinfo {author} {\bibfnamefont {E.}~\bibnamefont {Flurin}},
  \bibinfo {author} {\bibfnamefont {V.~V.}\ \bibnamefont {Ramasesh}}, \bibinfo
  {author} {\bibfnamefont {K.~B.}\ \bibnamefont {Whaley}},\ and\ \bibinfo
  {author} {\bibfnamefont {I.}~\bibnamefont {Siddiqi}},\ }\bibfield  {title}
  {\bibinfo {title} {Quantum dynamics of simultaneously measured non-commuting
  observables},\ }\href {https://doi.org/10.1038/nature19762} {\bibfield
  {journal} {\bibinfo  {journal} {Nature}\ }\textbf {\bibinfo {volume} {538}},\
  \bibinfo {pages} {491} (\bibinfo {year} {2016})}\BibitemShut {NoStop}%
\bibitem [{\citenamefont {Blais}\ \emph {et~al.}(2004)\citenamefont {Blais},
  \citenamefont {Huang}, \citenamefont {Wallraff}, \citenamefont {Girvin},\
  and\ \citenamefont {Schoelkopf}}]{Schoelkopf04}%
  \BibitemOpen
  \bibfield  {author} {\bibinfo {author} {\bibfnamefont {A.}~\bibnamefont
  {Blais}}, \bibinfo {author} {\bibfnamefont {R.-S.}\ \bibnamefont {Huang}},
  \bibinfo {author} {\bibfnamefont {A.}~\bibnamefont {Wallraff}}, \bibinfo
  {author} {\bibfnamefont {S.~M.}\ \bibnamefont {Girvin}},\ and\ \bibinfo
  {author} {\bibfnamefont {R.~J.}\ \bibnamefont {Schoelkopf}},\ }\bibfield
  {title} {\bibinfo {title} {Cavity quantum electrodynamics for superconducting
  electrical circuits: An architecture for quantum computation},\ }\href@noop
  {} {\bibfield  {journal} {\bibinfo  {journal} {Physical Review A}\ }\textbf
  {\bibinfo {volume} {69}},\ \bibinfo {pages} {062320} (\bibinfo {year}
  {2004})}\BibitemShut {NoStop}%
\bibitem [{\citenamefont {Mallet}\ \emph {et~al.}(2009)\citenamefont {Mallet},
  \citenamefont {Ong}, \citenamefont {Palacios-Laloy}, \citenamefont {Nguyen},
  \citenamefont {Bertet}, \citenamefont {Vion},\ and\ \citenamefont
  {Esteve}}]{Esteve09}%
  \BibitemOpen
  \bibfield  {author} {\bibinfo {author} {\bibfnamefont {F.}~\bibnamefont
  {Mallet}}, \bibinfo {author} {\bibfnamefont {F.~R.}\ \bibnamefont {Ong}},
  \bibinfo {author} {\bibfnamefont {A.}~\bibnamefont {Palacios-Laloy}},
  \bibinfo {author} {\bibfnamefont {F.}~\bibnamefont {Nguyen}}, \bibinfo
  {author} {\bibfnamefont {P.}~\bibnamefont {Bertet}}, \bibinfo {author}
  {\bibfnamefont {D.}~\bibnamefont {Vion}},\ and\ \bibinfo {author}
  {\bibfnamefont {D.}~\bibnamefont {Esteve}},\ }\bibfield  {title} {\bibinfo
  {title} {Single-shot qubit readout in circuit quantum electrodynamics},\
  }\href {https://doi.org/10.1038/nphys1400} {\bibfield  {journal} {\bibinfo
  {journal} {Nature Physics}\ }\textbf {\bibinfo {volume} {5}},\ \bibinfo
  {pages} {791} (\bibinfo {year} {2009})}\BibitemShut {NoStop}%
\bibitem [{\citenamefont {Walter}\ \emph {et~al.}(2017)\citenamefont {Walter},
  \citenamefont {Kurpiers}, \citenamefont {Gasparinetti}, \citenamefont
  {Magnard}, \citenamefont {Potočnik}, \citenamefont {Salathé}, \citenamefont
  {Pechal}, \citenamefont {Mondal}, \citenamefont {Oppliger}, \citenamefont
  {Eichler},\ and\ \citenamefont {Wallraff}}]{Wallraff17}%
  \BibitemOpen
  \bibfield  {author} {\bibinfo {author} {\bibfnamefont {T.}~\bibnamefont
  {Walter}}, \bibinfo {author} {\bibfnamefont {P.}~\bibnamefont {Kurpiers}},
  \bibinfo {author} {\bibfnamefont {S.}~\bibnamefont {Gasparinetti}}, \bibinfo
  {author} {\bibfnamefont {P.}~\bibnamefont {Magnard}}, \bibinfo {author}
  {\bibfnamefont {A.}~\bibnamefont {Potočnik}}, \bibinfo {author}
  {\bibfnamefont {Y.}~\bibnamefont {Salathé}}, \bibinfo {author}
  {\bibfnamefont {M.}~\bibnamefont {Pechal}}, \bibinfo {author} {\bibfnamefont
  {M.}~\bibnamefont {Mondal}}, \bibinfo {author} {\bibfnamefont
  {M.}~\bibnamefont {Oppliger}}, \bibinfo {author} {\bibfnamefont
  {C.}~\bibnamefont {Eichler}},\ and\ \bibinfo {author} {\bibfnamefont
  {A.}~\bibnamefont {Wallraff}},\ }\bibfield  {title} {\bibinfo {title} {Rapid
  high-fidelity single-shot dispersive readout of superconducting qubits},\
  }\href {https://doi.org/10.1103/PhysRevApplied.7.054020} {\bibfield
  {journal} {\bibinfo  {journal} {Physical Review Applied}\ }\textbf {\bibinfo
  {volume} {7}},\ \bibinfo {pages} {054020} (\bibinfo {year}
  {2017})}\BibitemShut {NoStop}%
\bibitem [{\citenamefont {Wang}\ \emph {et~al.}(2019)\citenamefont {Wang},
  \citenamefont {Miranowicz},\ and\ \citenamefont {Nori}}]{Nori19}%
  \BibitemOpen
  \bibfield  {author} {\bibinfo {author} {\bibfnamefont {X.}~\bibnamefont
  {Wang}}, \bibinfo {author} {\bibfnamefont {A.}~\bibnamefont {Miranowicz}},\
  and\ \bibinfo {author} {\bibfnamefont {F.}~\bibnamefont {Nori}},\ }\bibfield
  {title} {\bibinfo {title} {Ideal quantum nondemolition readout of a flux
  qubit without purcell limitations},\ }\href@noop {} {\bibfield  {journal}
  {\bibinfo  {journal} {Physical Review Applied}\ }\textbf {\bibinfo {volume}
  {12}} (\bibinfo {year} {2019})}\BibitemShut {NoStop}%
\bibitem [{\citenamefont {Clerk}\ \emph {et~al.}(2003)\citenamefont {Clerk},
  \citenamefont {Girvin},\ and\ \citenamefont {Stone}}]{Girvin03}%
  \BibitemOpen
  \bibfield  {author} {\bibinfo {author} {\bibfnamefont {A.}~\bibnamefont
  {Clerk}}, \bibinfo {author} {\bibfnamefont {S.}~\bibnamefont {Girvin}},\ and\
  \bibinfo {author} {\bibfnamefont {A.~D.}\ \bibnamefont {Stone}},\ }\bibfield
  {title} {\bibinfo {title} {Quantum-limited measurement and information in
  mesoscopic detectors},\ }\href@noop {} {\bibfield  {journal} {\bibinfo
  {journal} {Physical Review B}\ }\textbf {\bibinfo {volume} {67}},\ \bibinfo
  {pages} {165324} (\bibinfo {year} {2003})}\BibitemShut {NoStop}%
\bibitem [{\citenamefont {Clerk}\ \emph {et~al.}(2010)\citenamefont {Clerk},
  \citenamefont {Devoret}, \citenamefont {Girvin}, \citenamefont {Marquardt},\
  and\ \citenamefont {Schoelkopf}}]{Schoelkopf2010RPM}%
  \BibitemOpen
  \bibfield  {author} {\bibinfo {author} {\bibfnamefont {A.~A.}\ \bibnamefont
  {Clerk}}, \bibinfo {author} {\bibfnamefont {M.~H.}\ \bibnamefont {Devoret}},
  \bibinfo {author} {\bibfnamefont {S.~M.}\ \bibnamefont {Girvin}}, \bibinfo
  {author} {\bibfnamefont {F.}~\bibnamefont {Marquardt}},\ and\ \bibinfo
  {author} {\bibfnamefont {R.~J.}\ \bibnamefont {Schoelkopf}},\ }\bibfield
  {title} {\bibinfo {title} {Introduction to quantum noise, measurement, and
  amplification},\ }\href {https://doi.org/10.1103/RevModPhys.82.1155}
  {\bibfield  {journal} {\bibinfo  {journal} {Rev. Mod. Phys.}\ }\textbf
  {\bibinfo {volume} {82}},\ \bibinfo {pages} {1155} (\bibinfo {year}
  {2010})}\BibitemShut {NoStop}%
\bibitem [{\citenamefont {Blais}\ \emph {et~al.}(2021)\citenamefont {Blais},
  \citenamefont {Grimsmo}, \citenamefont {Girvin},\ and\ \citenamefont
  {Wallraff}}]{Wallraff2020review}%
  \BibitemOpen
  \bibfield  {author} {\bibinfo {author} {\bibfnamefont {A.}~\bibnamefont
  {Blais}}, \bibinfo {author} {\bibfnamefont {A.~L.}\ \bibnamefont {Grimsmo}},
  \bibinfo {author} {\bibfnamefont {S.~M.}\ \bibnamefont {Girvin}},\ and\
  \bibinfo {author} {\bibfnamefont {A.}~\bibnamefont {Wallraff}},\ }\bibfield
  {title} {\bibinfo {title} {Circuit quantum electrodynamics},\ }\href@noop {}
  {\bibfield  {journal} {\bibinfo  {journal} {Reviews of Modern Physics}\
  }\textbf {\bibinfo {volume} {93}} (\bibinfo {year} {2021})}\BibitemShut
  {NoStop}%
\bibitem [{\citenamefont {Ikonen}\ \emph {et~al.}(2019)\citenamefont {Ikonen},
  \citenamefont {Goetz}, \citenamefont {Ilves}, \citenamefont {Keranen},
  \citenamefont {Gunyho}, \citenamefont {Partanen}, \citenamefont {Tan},
  \citenamefont {Hazra}, \citenamefont {Gronberg}, \citenamefont {Vesterinen},
  \citenamefont {Simbierowicz}, \citenamefont {Hassel},\ and\ \citenamefont
  {Möttönen}}]{Mottonen19}%
  \BibitemOpen
  \bibfield  {author} {\bibinfo {author} {\bibfnamefont {J.}~\bibnamefont
  {Ikonen}}, \bibinfo {author} {\bibfnamefont {J.}~\bibnamefont {Goetz}},
  \bibinfo {author} {\bibfnamefont {J.}~\bibnamefont {Ilves}}, \bibinfo
  {author} {\bibfnamefont {A.}~\bibnamefont {Keranen}}, \bibinfo {author}
  {\bibfnamefont {A.~M.}\ \bibnamefont {Gunyho}}, \bibinfo {author}
  {\bibfnamefont {M.}~\bibnamefont {Partanen}}, \bibinfo {author}
  {\bibfnamefont {K.~Y.}\ \bibnamefont {Tan}}, \bibinfo {author} {\bibfnamefont
  {D.}~\bibnamefont {Hazra}}, \bibinfo {author} {\bibfnamefont
  {L.}~\bibnamefont {Gronberg}}, \bibinfo {author} {\bibfnamefont
  {V.}~\bibnamefont {Vesterinen}}, \bibinfo {author} {\bibfnamefont
  {S.}~\bibnamefont {Simbierowicz}}, \bibinfo {author} {\bibfnamefont
  {J.}~\bibnamefont {Hassel}},\ and\ \bibinfo {author} {\bibfnamefont
  {M.}~\bibnamefont {Möttönen}},\ }\bibfield  {title} {\bibinfo {title}
  {Qubit measurement by multichannel driving},\ }\href
  {https://www.ncbi.nlm.nih.gov/pubmed/30932559} {\bibfield  {journal}
  {\bibinfo  {journal} {Physical Review Letters}\ }\textbf {\bibinfo {volume}
  {122}},\ \bibinfo {pages} {080503} (\bibinfo {year} {2019})}\BibitemShut
  {NoStop}%
\bibitem [{\citenamefont {Touzard}\ \emph {et~al.}(2019)\citenamefont
  {Touzard}, \citenamefont {Kou}, \citenamefont {Frattini}, \citenamefont
  {Sivak}, \citenamefont {Puri}, \citenamefont {Grimm}, \citenamefont
  {Frunzio}, \citenamefont {Shankar},\ and\ \citenamefont
  {Devoret}}]{Devoret19}%
  \BibitemOpen
  \bibfield  {author} {\bibinfo {author} {\bibfnamefont {S.}~\bibnamefont
  {Touzard}}, \bibinfo {author} {\bibfnamefont {A.}~\bibnamefont {Kou}},
  \bibinfo {author} {\bibfnamefont {N.~E.}\ \bibnamefont {Frattini}}, \bibinfo
  {author} {\bibfnamefont {V.~V.}\ \bibnamefont {Sivak}}, \bibinfo {author}
  {\bibfnamefont {S.}~\bibnamefont {Puri}}, \bibinfo {author} {\bibfnamefont
  {A.}~\bibnamefont {Grimm}}, \bibinfo {author} {\bibfnamefont
  {L.}~\bibnamefont {Frunzio}}, \bibinfo {author} {\bibfnamefont
  {S.}~\bibnamefont {Shankar}},\ and\ \bibinfo {author} {\bibfnamefont {M.~H.}\
  \bibnamefont {Devoret}},\ }\bibfield  {title} {\bibinfo {title} {Gated
  conditional displacement readout of superconducting qubits},\ }\href
  {https://doi.org/10.1103/PhysRevLett.122.080502} {\bibfield  {journal}
  {\bibinfo  {journal} {Physical Review Letters}\ }\textbf {\bibinfo {volume}
  {122}},\ \bibinfo {pages} {080502} (\bibinfo {year} {2019})}\BibitemShut
  {NoStop}%
\bibitem [{\citenamefont {Reed}\ \emph {et~al.}(2010)\citenamefont {Reed},
  \citenamefont {DiCarlo}, \citenamefont {Johnson}, \citenamefont {Sun},
  \citenamefont {Schuster}, \citenamefont {Frunzio},\ and\ \citenamefont
  {Schoelkopf}}]{Schoelkopf2010}%
  \BibitemOpen
  \bibfield  {author} {\bibinfo {author} {\bibfnamefont {M.~D.}\ \bibnamefont
  {Reed}}, \bibinfo {author} {\bibfnamefont {L.}~\bibnamefont {DiCarlo}},
  \bibinfo {author} {\bibfnamefont {B.~R.}\ \bibnamefont {Johnson}}, \bibinfo
  {author} {\bibfnamefont {L.}~\bibnamefont {Sun}}, \bibinfo {author}
  {\bibfnamefont {D.~I.}\ \bibnamefont {Schuster}}, \bibinfo {author}
  {\bibfnamefont {L.}~\bibnamefont {Frunzio}},\ and\ \bibinfo {author}
  {\bibfnamefont {R.~J.}\ \bibnamefont {Schoelkopf}},\ }\bibfield  {title}
  {\bibinfo {title} {High-fidelity readout in circuit quantum electrodynamics
  using the jaynes-cummings nonlinearity},\ }\href
  {https://doi.org/10.1103/PhysRevLett.105.173601} {\bibfield  {journal}
  {\bibinfo  {journal} {Physical Review Letters}\ }\textbf {\bibinfo {volume}
  {105}},\ \bibinfo {pages} {173601} (\bibinfo {year} {2010})}\BibitemShut
  {NoStop}%
\bibitem [{\citenamefont {Boissonneault}\ \emph {et~al.}(2010)\citenamefont
  {Boissonneault}, \citenamefont {Gambetta},\ and\ \citenamefont
  {Blais}}]{Blais10}%
  \BibitemOpen
  \bibfield  {author} {\bibinfo {author} {\bibfnamefont {M.}~\bibnamefont
  {Boissonneault}}, \bibinfo {author} {\bibfnamefont {J.~M.}\ \bibnamefont
  {Gambetta}},\ and\ \bibinfo {author} {\bibfnamefont {A.}~\bibnamefont
  {Blais}},\ }\bibfield  {title} {\bibinfo {title} {Improved superconducting
  qubit readout by qubit-induced nonlinearities},\ }\href
  {https://doi.org/10.1103/PhysRevLett.105.100504} {\bibfield  {journal}
  {\bibinfo  {journal} {Physical Review Letters}\ }\textbf {\bibinfo {volume}
  {105}},\ \bibinfo {pages} {100504} (\bibinfo {year} {2010})}\BibitemShut
  {NoStop}%
\bibitem [{\citenamefont {Gao}\ \emph {et~al.}(2008)\citenamefont {Gao},
  \citenamefont {Zmuidzinas}, \citenamefont {of~Technology. Division~of
  Engineering},\ and\ \citenamefont {Science}}]{gao2008physics}%
  \BibitemOpen
  \bibfield  {author} {\bibinfo {author} {\bibfnamefont {J.}~\bibnamefont
  {Gao}}, \bibinfo {author} {\bibfnamefont {J.}~\bibnamefont {Zmuidzinas}},
  \bibinfo {author} {\bibfnamefont {C.~I.}\ \bibnamefont {of~Technology.
  Division~of Engineering}},\ and\ \bibinfo {author} {\bibfnamefont
  {A.}~\bibnamefont {Science}},\ }\href
  {https://books.google.com/books?id=Tbg3QwAACAAJ} {\emph {\bibinfo {title}
  {The Physics of Superconducting Microwave Resonators}}},\ CIT theses\
  (\bibinfo  {publisher} {California Institute of Technology},\ \bibinfo {year}
  {2008})\BibitemShut {NoStop}%
\bibitem [{\citenamefont {Bradley}(2018)}]{Bradley2018}%
  \BibitemOpen
  \bibfield  {author} {\bibinfo {author} {\bibfnamefont {R.}~\bibnamefont
  {Bradley}},\ }\bibinfo {title} {Modification of a commercial phase shifter
  for cryogenic applications: Proceedings of the 2nd international workshop}\
  (\bibinfo {year} {2018})\ pp.\ \bibinfo {pages} {39--44}\BibitemShut
  {NoStop}%
\bibitem [{\citenamefont {Kokkoniemi}\ \emph {et~al.}(2017)\citenamefont
  {Kokkoniemi}, \citenamefont {Ollikainen}, \citenamefont {Lake}, \citenamefont
  {Saarenpää}, \citenamefont {Tan}, \citenamefont {Kokkala}, \citenamefont
  {Dağ}, \citenamefont {Govenius},\ and\ \citenamefont
  {Möttönen}}]{Mikko2017}%
  \BibitemOpen
  \bibfield  {author} {\bibinfo {author} {\bibfnamefont {R.}~\bibnamefont
  {Kokkoniemi}}, \bibinfo {author} {\bibfnamefont {T.}~\bibnamefont
  {Ollikainen}}, \bibinfo {author} {\bibfnamefont {R.~E.}\ \bibnamefont
  {Lake}}, \bibinfo {author} {\bibfnamefont {S.}~\bibnamefont {Saarenpää}},
  \bibinfo {author} {\bibfnamefont {K.~Y.}\ \bibnamefont {Tan}}, \bibinfo
  {author} {\bibfnamefont {J.~I.}\ \bibnamefont {Kokkala}}, \bibinfo {author}
  {\bibfnamefont {C.~B.}\ \bibnamefont {Dağ}}, \bibinfo {author}
  {\bibfnamefont {J.}~\bibnamefont {Govenius}},\ and\ \bibinfo {author}
  {\bibfnamefont {M.}~\bibnamefont {Möttönen}},\ }\bibfield  {title}
  {\bibinfo {title} {Flux-tunable phase shifter for microwaves},\ }\href
  {https://doi.org/10.1038/s41598-017-15190-2} {\bibfield  {journal} {\bibinfo
  {journal} {Scientific Reports}\ }\textbf {\bibinfo {volume} {7}},\ \bibinfo
  {pages} {14713} (\bibinfo {year} {2017})}\BibitemShut {NoStop}%
\bibitem [{\citenamefont {Zhang}\ \emph {et~al.}(2020)\citenamefont {Zhang},
  \citenamefont {Li}, \citenamefont {Kokkoniemi}, \citenamefont {Yan},
  \citenamefont {Liu}, \citenamefont {Partanen}, \citenamefont {Tan},
  \citenamefont {He}, \citenamefont {Ji}, \citenamefont {Grönberg},\ and\
  \citenamefont {Möttönen}}]{Mikko2020}%
  \BibitemOpen
  \bibfield  {author} {\bibinfo {author} {\bibfnamefont {J.}~\bibnamefont
  {Zhang}}, \bibinfo {author} {\bibfnamefont {T.}~\bibnamefont {Li}}, \bibinfo
  {author} {\bibfnamefont {R.}~\bibnamefont {Kokkoniemi}}, \bibinfo {author}
  {\bibfnamefont {C.}~\bibnamefont {Yan}}, \bibinfo {author} {\bibfnamefont
  {W.}~\bibnamefont {Liu}}, \bibinfo {author} {\bibfnamefont {M.}~\bibnamefont
  {Partanen}}, \bibinfo {author} {\bibfnamefont {K.~Y.}\ \bibnamefont {Tan}},
  \bibinfo {author} {\bibfnamefont {M.}~\bibnamefont {He}}, \bibinfo {author}
  {\bibfnamefont {L.}~\bibnamefont {Ji}}, \bibinfo {author} {\bibfnamefont
  {L.}~\bibnamefont {Grönberg}},\ and\ \bibinfo {author} {\bibfnamefont
  {M.}~\bibnamefont {Möttönen}},\ }\bibfield  {title} {\bibinfo {title}
  {Broadband tunable phase shifter for microwaves},\ }\href
  {https://doi.org/10.1063/5.0006499} {\bibfield  {journal} {\bibinfo
  {journal} {AIP Advances}\ }\textbf {\bibinfo {volume} {10}},\ \bibinfo
  {pages} {065128} (\bibinfo {year} {2020})}\BibitemShut {NoStop}%
\bibitem [{\citenamefont {Naaman}\ \emph {et~al.}(2017)\citenamefont {Naaman},
  \citenamefont {Strong}, \citenamefont {Ferguson}, \citenamefont {Egan},
  \citenamefont {Bailey},\ and\ \citenamefont {Hinkey}}]{Naaman2017}%
  \BibitemOpen
  \bibfield  {author} {\bibinfo {author} {\bibfnamefont {O.}~\bibnamefont
  {Naaman}}, \bibinfo {author} {\bibfnamefont {J.~A.}\ \bibnamefont {Strong}},
  \bibinfo {author} {\bibfnamefont {D.~G.}\ \bibnamefont {Ferguson}}, \bibinfo
  {author} {\bibfnamefont {J.}~\bibnamefont {Egan}}, \bibinfo {author}
  {\bibfnamefont {N.}~\bibnamefont {Bailey}},\ and\ \bibinfo {author}
  {\bibfnamefont {R.~T.}\ \bibnamefont {Hinkey}},\ }\bibfield  {title}
  {\bibinfo {title} {Josephson junction microwave modulators for qubit
  control},\ }\href {https://doi.org/10.1063/1.4976809} {\bibfield  {journal}
  {\bibinfo  {journal} {Journal of Applied Physics}\ }\textbf {\bibinfo
  {volume} {121}},\ \bibinfo {pages} {073904} (\bibinfo {year}
  {2017})}\BibitemShut {NoStop}%
\bibitem [{\citenamefont {Eder}\ \emph {et~al.}(2018)\citenamefont {Eder},
  \citenamefont {Ramos}, \citenamefont {Goetz}, \citenamefont {Fischer},
  \citenamefont {Pogorzalek}, \citenamefont {Martínez}, \citenamefont
  {Menzel}, \citenamefont {Loacker}, \citenamefont {Xie}, \citenamefont
  {Garcia-Ripoll}, \citenamefont {Fedorov}, \citenamefont {Marx}, \citenamefont
  {Deppe},\ and\ \citenamefont {Gross}}]{Gross2018}%
  \BibitemOpen
  \bibfield  {author} {\bibinfo {author} {\bibfnamefont {P.}~\bibnamefont
  {Eder}}, \bibinfo {author} {\bibfnamefont {T.}~\bibnamefont {Ramos}},
  \bibinfo {author} {\bibfnamefont {J.}~\bibnamefont {Goetz}}, \bibinfo
  {author} {\bibfnamefont {M.}~\bibnamefont {Fischer}}, \bibinfo {author}
  {\bibfnamefont {S.}~\bibnamefont {Pogorzalek}}, \bibinfo {author}
  {\bibfnamefont {J.~P.}\ \bibnamefont {Martínez}}, \bibinfo {author}
  {\bibfnamefont {E.~P.}\ \bibnamefont {Menzel}}, \bibinfo {author}
  {\bibfnamefont {F.}~\bibnamefont {Loacker}}, \bibinfo {author} {\bibfnamefont
  {E.}~\bibnamefont {Xie}}, \bibinfo {author} {\bibfnamefont {J.~J.}\
  \bibnamefont {Garcia-Ripoll}}, \bibinfo {author} {\bibfnamefont {K.~G.}\
  \bibnamefont {Fedorov}}, \bibinfo {author} {\bibfnamefont {A.}~\bibnamefont
  {Marx}}, \bibinfo {author} {\bibfnamefont {F.}~\bibnamefont {Deppe}},\ and\
  \bibinfo {author} {\bibfnamefont {R.}~\bibnamefont {Gross}},\ }\bibfield
  {title} {\bibinfo {title} {Quantum probe of an on-chip broadband
  interferometer for quantum microwave photonics},\ }\href
  {https://doi.org/10.1088/1361-6668/aad8f4} {\bibfield  {journal} {\bibinfo
  {journal} {Superconductor Science and Technology}\ }\textbf {\bibinfo
  {volume} {31}},\ \bibinfo {pages} {115002} (\bibinfo {year}
  {2018})}\BibitemShut {NoStop}%
\bibitem [{\citenamefont {Pogorzalek}\ \emph {et~al.}(2019)\citenamefont
  {Pogorzalek}, \citenamefont {Fedorov}, \citenamefont {Xu}, \citenamefont
  {Parra-Rodriguez}, \citenamefont {Sanz}, \citenamefont {Fischer},
  \citenamefont {Xie}, \citenamefont {Inomata}, \citenamefont {Nakamura},
  \citenamefont {Solano}, \citenamefont {Marx}, \citenamefont {Deppe},\ and\
  \citenamefont {Gross}}]{Gross2019}%
  \BibitemOpen
  \bibfield  {author} {\bibinfo {author} {\bibfnamefont {S.}~\bibnamefont
  {Pogorzalek}}, \bibinfo {author} {\bibfnamefont {K.~G.}\ \bibnamefont
  {Fedorov}}, \bibinfo {author} {\bibfnamefont {M.}~\bibnamefont {Xu}},
  \bibinfo {author} {\bibfnamefont {A.}~\bibnamefont {Parra-Rodriguez}},
  \bibinfo {author} {\bibfnamefont {M.}~\bibnamefont {Sanz}}, \bibinfo {author}
  {\bibfnamefont {M.}~\bibnamefont {Fischer}}, \bibinfo {author} {\bibfnamefont
  {E.}~\bibnamefont {Xie}}, \bibinfo {author} {\bibfnamefont {K.}~\bibnamefont
  {Inomata}}, \bibinfo {author} {\bibfnamefont {Y.}~\bibnamefont {Nakamura}},
  \bibinfo {author} {\bibfnamefont {E.}~\bibnamefont {Solano}}, \bibinfo
  {author} {\bibfnamefont {A.}~\bibnamefont {Marx}}, \bibinfo {author}
  {\bibfnamefont {F.}~\bibnamefont {Deppe}},\ and\ \bibinfo {author}
  {\bibfnamefont {R.}~\bibnamefont {Gross}},\ }\bibfield  {title} {\bibinfo
  {title} {Secure quantum remote state preparation of squeezed microwave
  states},\ }\href {https://doi.org/10.1038/s41467-019-10727-7} {\bibfield
  {journal} {\bibinfo  {journal} {Nature Communications}\ }\textbf {\bibinfo
  {volume} {10}},\ \bibinfo {pages} {2604} (\bibinfo {year}
  {2019})}\BibitemShut {NoStop}%
\bibitem [{\citenamefont {Probst}\ \emph {et~al.}(2015)\citenamefont {Probst},
  \citenamefont {Song}, \citenamefont {Bushev}, \citenamefont {Ustinov},\ and\
  \citenamefont {Weides}}]{Weides2015}%
  \BibitemOpen
  \bibfield  {author} {\bibinfo {author} {\bibfnamefont {S.}~\bibnamefont
  {Probst}}, \bibinfo {author} {\bibfnamefont {F.~B.}\ \bibnamefont {Song}},
  \bibinfo {author} {\bibfnamefont {P.~A.}\ \bibnamefont {Bushev}}, \bibinfo
  {author} {\bibfnamefont {A.~V.}\ \bibnamefont {Ustinov}},\ and\ \bibinfo
  {author} {\bibfnamefont {M.}~\bibnamefont {Weides}},\ }\bibfield  {title}
  {\bibinfo {title} {Efficient and robust analysis of complex scattering data
  under noise in microwave resonators},\ }\href@noop {} {\bibfield  {journal}
  {\bibinfo  {journal} {Review of Scientific Instruments}\ }\textbf {\bibinfo
  {volume} {86}},\ \bibinfo {pages} {024706} (\bibinfo {year}
  {2015})}\BibitemShut {NoStop}%
\bibitem [{\citenamefont {Krantz}\ \emph {et~al.}(2016)\citenamefont {Krantz},
  \citenamefont {Bengtsson}, \citenamefont {Simoen}, \citenamefont
  {Gustavsson}, \citenamefont {Shumeiko}, \citenamefont {Oliver}, \citenamefont
  {Wilson}, \citenamefont {Delsing},\ and\ \citenamefont
  {Bylander}}]{Delsing16}%
  \BibitemOpen
  \bibfield  {author} {\bibinfo {author} {\bibfnamefont {P.}~\bibnamefont
  {Krantz}}, \bibinfo {author} {\bibfnamefont {A.}~\bibnamefont {Bengtsson}},
  \bibinfo {author} {\bibfnamefont {M.}~\bibnamefont {Simoen}}, \bibinfo
  {author} {\bibfnamefont {S.}~\bibnamefont {Gustavsson}}, \bibinfo {author}
  {\bibfnamefont {V.}~\bibnamefont {Shumeiko}}, \bibinfo {author}
  {\bibfnamefont {W.~D.}\ \bibnamefont {Oliver}}, \bibinfo {author}
  {\bibfnamefont {C.~M.}\ \bibnamefont {Wilson}}, \bibinfo {author}
  {\bibfnamefont {P.}~\bibnamefont {Delsing}},\ and\ \bibinfo {author}
  {\bibfnamefont {J.}~\bibnamefont {Bylander}},\ }\bibfield  {title} {\bibinfo
  {title} {Single-shot read-out of a superconducting qubit using a josephson
  parametric oscillator},\ }\href {https://doi.org/10.1038/ncomms11417}
  {\bibfield  {journal} {\bibinfo  {journal} {Nature Communications}\ }\textbf
  {\bibinfo {volume} {7}},\ \bibinfo {pages} {11417} (\bibinfo {year}
  {2016})}\BibitemShut {NoStop}%
\bibitem [{\citenamefont {Krantz}\ \emph {et~al.}(2019)\citenamefont {Krantz},
  \citenamefont {Kjaergaard}, \citenamefont {Yan}, \citenamefont {Orlando},
  \citenamefont {Gustavsson},\ and\ \citenamefont {Oliver}}]{Oliver2019}%
  \BibitemOpen
  \bibfield  {author} {\bibinfo {author} {\bibfnamefont {P.}~\bibnamefont
  {Krantz}}, \bibinfo {author} {\bibfnamefont {M.}~\bibnamefont {Kjaergaard}},
  \bibinfo {author} {\bibfnamefont {F.}~\bibnamefont {Yan}}, \bibinfo {author}
  {\bibfnamefont {T.~P.}\ \bibnamefont {Orlando}}, \bibinfo {author}
  {\bibfnamefont {S.}~\bibnamefont {Gustavsson}},\ and\ \bibinfo {author}
  {\bibfnamefont {W.~D.}\ \bibnamefont {Oliver}},\ }\bibfield  {title}
  {\bibinfo {title} {A quantum engineer's guide to superconducting qubits},\
  }\href {https://doi.org/10.1063/1.5089550} {\bibfield  {journal} {\bibinfo
  {journal} {Applied Physics Reviews}\ }\textbf {\bibinfo {volume} {6}},\
  \bibinfo {pages} {021318} (\bibinfo {year} {2019})}\BibitemShut {NoStop}%
\bibitem [{\citenamefont {Kjaergaard}\ \emph {et~al.}(2020)\citenamefont
  {Kjaergaard}, \citenamefont {Schwartz}, \citenamefont {Braumüller},
  \citenamefont {Krantz}, \citenamefont {Wang}, \citenamefont {Gustavsson},\
  and\ \citenamefont {Oliver}}]{Oliver2020}%
  \BibitemOpen
  \bibfield  {author} {\bibinfo {author} {\bibfnamefont {M.}~\bibnamefont
  {Kjaergaard}}, \bibinfo {author} {\bibfnamefont {M.~E.}\ \bibnamefont
  {Schwartz}}, \bibinfo {author} {\bibfnamefont {J.}~\bibnamefont
  {Braumüller}}, \bibinfo {author} {\bibfnamefont {P.}~\bibnamefont {Krantz}},
  \bibinfo {author} {\bibfnamefont {J.~I.~J.}\ \bibnamefont {Wang}}, \bibinfo
  {author} {\bibfnamefont {S.}~\bibnamefont {Gustavsson}},\ and\ \bibinfo
  {author} {\bibfnamefont {W.~D.}\ \bibnamefont {Oliver}},\ }\bibfield  {title}
  {\bibinfo {title} {Superconducting qubits: Current state of play},\ }\href
  {https://doi.org/10.1146/annurev-conmatphys-031119-050605} {\bibfield
  {journal} {\bibinfo  {journal} {Annual Review of Condensed Matter Physics}\
  }\textbf {\bibinfo {volume} {11}},\ \bibinfo {pages} {369} (\bibinfo {year}
  {2020})}\BibitemShut {NoStop}%
\bibitem [{\citenamefont {Place}\ \emph {et~al.}(2021)\citenamefont {Place},
  \citenamefont {Rodgers}, \citenamefont {Mundada}, \citenamefont {Smitham},
  \citenamefont {Fitzpatrick}, \citenamefont {Leng}, \citenamefont {Premkumar},
  \citenamefont {Bryon}, \citenamefont {Vrajitoarea}, \citenamefont {Sussman},\
  and\ \citenamefont {et~al.}}]{Houck2020}%
  \BibitemOpen
  \bibfield  {author} {\bibinfo {author} {\bibfnamefont {A.~P.~M.}\
  \bibnamefont {Place}}, \bibinfo {author} {\bibfnamefont {L.~V.~H.}\
  \bibnamefont {Rodgers}}, \bibinfo {author} {\bibfnamefont {P.}~\bibnamefont
  {Mundada}}, \bibinfo {author} {\bibfnamefont {B.~M.}\ \bibnamefont
  {Smitham}}, \bibinfo {author} {\bibfnamefont {M.}~\bibnamefont
  {Fitzpatrick}}, \bibinfo {author} {\bibfnamefont {Z.}~\bibnamefont {Leng}},
  \bibinfo {author} {\bibfnamefont {A.}~\bibnamefont {Premkumar}}, \bibinfo
  {author} {\bibfnamefont {J.}~\bibnamefont {Bryon}}, \bibinfo {author}
  {\bibfnamefont {A.}~\bibnamefont {Vrajitoarea}}, \bibinfo {author}
  {\bibfnamefont {S.}~\bibnamefont {Sussman}},\ and\ \bibinfo {author}
  {\bibnamefont {et~al.}},\ }\bibfield  {title} {\bibinfo {title} {New material
  platform for superconducting transmon qubits with coherence times exceeding
  0.3 milliseconds},\ }\href@noop {} {\bibfield  {journal} {\bibinfo  {journal}
  {Nature Communications}\ }\textbf {\bibinfo {volume} {12}} (\bibinfo {year}
  {2021})}\BibitemShut {NoStop}%
\end{thebibliography}%


\begin{thebibliography}{4}
\expandafter\ifx\csname natexlab\endcsname\relax\def\natexlab#1{#1}\fi
\expandafter\ifx\csname bibnamefont\endcsname\relax
  \def\bibnamefont#1{#1}\fi
\expandafter\ifx\csname bibfnamefont\endcsname\relax
  \def\bibfnamefont#1{#1}\fi
\expandafter\ifx\csname citenamefont\endcsname\relax
  \def\citenamefont#1{#1}\fi
\expandafter\ifx\csname url\endcsname\relax
  \def\url#1{\texttt{#1}}\fi
\expandafter\ifx\csname urlprefix\endcsname\relax\def\urlprefix{URL }\fi
\providecommand{\bibinfo}[2]{#2}
\providecommand{\eprint}[2][]{\url{#2}}

\bibitem[{\citenamefont{Eichler}(2013)}]{Eichlerthesis}
\bibinfo{author}{\bibfnamefont{C.}~\bibnamefont{Eichler}}, Ph.D. thesis,
  \bibinfo{school}{ETH Zurich}, \bibinfo{address}{Zurich}
  (\bibinfo{year}{2013}).

\bibitem[{\citenamefont{Sete et~al.}(2015)\citenamefont{Sete, Martinis, and
  Korotkov}}]{Martinis15}
\bibinfo{author}{\bibfnamefont{E.~A.} \bibnamefont{Sete}},
  \bibinfo{author}{\bibfnamefont{J.~M.} \bibnamefont{Martinis}},
  \bibnamefont{and} \bibinfo{author}{\bibfnamefont{A.~N.}
  \bibnamefont{Korotkov}}, \bibinfo{journal}{Physical Review A}
  \textbf{\bibinfo{volume}{92}}, \bibinfo{pages}{012325}
  (\bibinfo{year}{2015}).

\bibitem[{\citenamefont{Krantz et~al.}(2016)\citenamefont{Krantz, Bengtsson,
  Simoen, Gustavsson, Shumeiko, Oliver, Wilson, Delsing, and
  Bylander}}]{Delsing16}
\bibinfo{author}{\bibfnamefont{P.}~\bibnamefont{Krantz}},
  \bibinfo{author}{\bibfnamefont{A.}~\bibnamefont{Bengtsson}},
  \bibinfo{author}{\bibfnamefont{M.}~\bibnamefont{Simoen}},
  \bibinfo{author}{\bibfnamefont{S.}~\bibnamefont{Gustavsson}},
  \bibinfo{author}{\bibfnamefont{V.}~\bibnamefont{Shumeiko}},
  \bibinfo{author}{\bibfnamefont{W.~D.} \bibnamefont{Oliver}},
  \bibinfo{author}{\bibfnamefont{C.~M.} \bibnamefont{Wilson}},
  \bibinfo{author}{\bibfnamefont{P.}~\bibnamefont{Delsing}}, \bibnamefont{and}
  \bibinfo{author}{\bibfnamefont{J.}~\bibnamefont{Bylander}},
  \bibinfo{journal}{Nature Communications} \textbf{\bibinfo{volume}{7}},
  \bibinfo{pages}{11417} (\bibinfo{year}{2016}), ISSN
  \bibinfo{issn}{2041-1723}.

\bibitem[{\citenamefont{Krinner et~al.}(2019)\citenamefont{Krinner, Storz,
  Kurpiers, Magnard, Heinsoo, Keller, Lütolf, Eichler, and
  Wallraff}}]{Wallrafffridge2019}
\bibinfo{author}{\bibfnamefont{S.}~\bibnamefont{Krinner}},
  \bibinfo{author}{\bibfnamefont{S.}~\bibnamefont{Storz}},
  \bibinfo{author}{\bibfnamefont{P.}~\bibnamefont{Kurpiers}},
  \bibinfo{author}{\bibfnamefont{P.}~\bibnamefont{Magnard}},
  \bibinfo{author}{\bibfnamefont{J.}~\bibnamefont{Heinsoo}},
  \bibinfo{author}{\bibfnamefont{R.}~\bibnamefont{Keller}},
  \bibinfo{author}{\bibfnamefont{J.}~\bibnamefont{Lütolf}},
  \bibinfo{author}{\bibfnamefont{C.}~\bibnamefont{Eichler}}, \bibnamefont{and}
  \bibinfo{author}{\bibfnamefont{A.}~\bibnamefont{Wallraff}},
  \bibinfo{journal}{EPJ Quantum Technology} \textbf{\bibinfo{volume}{6}},
  \bibinfo{pages}{2} (\bibinfo{year}{2019}), ISSN \bibinfo{issn}{2196-0763}.

\end{thebibliography}

\end{document}


\newcommand{\dif}{\mathrm{d}}

\title{Supplementary Information for ``Improved superconducting qubit state readout by path interference''}

\author{Zhiling Wang}
\altaffiliation{These two authors contributed equally to this work.}

\author{Zenghui Bao}
\altaffiliation{These two authors contributed equally to this work.}

\author{Yukai Wu}

\author{Yan Li}

\author{Cheng Ma}

\author{Tianqi Cai}

\author{Yipu Song}

\author{Hongyi Zhang}
\email{hyzhang2016@tsinghua.edu.cn}

\author{Luming Duan}
\email{lmduan@tsinghua.edu.cn}

\affiliation{Center for Quantum Information, Institute for Interdisciplinary Information Sciences, Tsinghua University, Beijing 100084, PR China}

\date{\today}

\maketitle

\section{\label{sec:theo}Theory}

For a superconducting qubit dispersively coupled to a cavity, the Hamiltonian of the system can be written as $H=(\omega_r+\chi\sigma_z)a^\dagger a+\omega_q \sigma_z/2$, which means the resonant frequency of the cavity depends on the state of the qubit. The photon state in the readout cavity can be calculated from the evolution of annihilation operator,
\begin{equation} 
\frac{\partial \alpha}{\partial t}=-i(\omega_d-(\omega_r+\chi\sigma_z))\alpha-\frac{\kappa}{2} \alpha+\sqrt{\kappa_c/2}\alpha_{in}
\label{intraalpha}
\end{equation}
where $\kappa_c$ is the damping rate of the cavity to the transmission line and $\kappa=\kappa_c+\kappa_i$ is the total damping rate, with $\kappa_i$ being the internal damping rate of the cavity. $\omega_d$ is the frequency of the probe field and $\alpha_{in}$ represents the strength of the probe field.
From Eq.~(\ref{intraalpha}), the steady state in the readout cavity can be written as $\alpha=\frac{\sqrt{\kappa_c/2}\alpha_{in}}{i(\omega_d-(\omega_r+\chi\sigma_z))+\kappa/2}$. According to the input-output theory, photon states of transmission and reflection from the hanger cavity can be written as
\begin{equation}
\begin{split}
\alpha_T(\omega_d,\sigma_z)&=\alpha_{in}-\sqrt{\kappa_c/2}\alpha\\
&=(1-\frac{\kappa_c/\kappa}{1+2i(\omega_d-\omega_r-\chi\sigma_z)/\kappa})\alpha_{in},\\
\alpha_R(\omega_d,\sigma_z)&=-\sqrt{\kappa_c/2}\alpha\\
&=-\frac{\kappa_c/\kappa}{1+2i(\omega_d-\omega_r-\chi\sigma_z)/\kappa}\alpha_{in}
\end{split}
\label{ioput}
\end{equation}
Therefore, for a given qubit state, when sweeping the probe frequency the measured cavity response would be a circle on the phase plane, which is referred to IQ circle in the following context. Examples can be found in Fig. 2(b) in the main text.

For a given probe frequency, the distance between the two pointer states corresponding to the qubit states $\ket{g}$ and $\ket{e}$ from $T$ and $R$ would be 
\begin{equation}
\begin{split}
D_{ge}^{T}&=|\alpha_T(\omega_d,1)-\alpha_T(\omega_d,-1)|\\
&=\frac{4\kappa_c\chi|\alpha_{in}|}{\sqrt{(\kappa^2+4\chi^2-4(\omega_d-\omega_r)^2)^2+16\kappa^2(\omega_d-\omega_r)^2}}\\
D_{ge}^{R}&=|\alpha_R(\omega_d,1)-\alpha_R(\omega_d,-1)|\\
&=\frac{4\kappa_c\chi|\alpha_{in}|}{\sqrt{(\kappa^2+4\chi^2-4(\omega_d-\omega_r)^2)^2+16\kappa^2(\omega_d-\omega_r)^2}}
\label{tr-dis}
\end{split}
\end{equation}
From Eq.~(\ref{tr-dis}), one can see that the distance between the two pointer states are equal, as a result of the symmetrical coupling of the hanger cavity. 

In order to effectively collect the cavity photons, we use a beam splitter to combine transmission and reflection signals, generating interference signals labeled as plus mode and minus mode,
\begin{equation}
\alpha^{int}(\omega_d,\sigma_z)= \frac{\alpha_{T}(\omega_d,\sigma_z) \pm e^{i\theta_{RT}}\alpha_{R}(\omega_d,\sigma_z)}{\sqrt{2}}
\label{alpha-int}
\end{equation}
where $\theta_{RT}$ characterizes the phase difference between transmission mode and reflection mode, due to the imbalanced circuit lengths when they are interfered at the beam splitter. $\theta_{RT}$ can be measured in the experiment.
The distance between the two pointer states for the interference output would be

\begin{equation}
\begin{split}
&D_{ge}^{int}(\theta)=|\alpha_{int}(\omega_d,1)\pm \alpha_{int}(\omega_d,-1)|\\
&=\left|\frac{1\pm e^{i\theta_{rt}}}{\sqrt{2}}\right| \frac{4\kappa_c\chi|\alpha_{in}|}{\sqrt{(\kappa^2+4\chi^2-4(\omega_d-\omega_r)^2)^2+16\kappa^2(\omega_d-\omega_r)^2}}
\end{split}
\end{equation}
One can see that the distance between the two pointer states from interference output can be maximally enlarged by a factor of $\sqrt{2}$ compared with that from either transmission or reflection, as illustrated in inset of Fig.~3 in the main text. 

\begin{figure*}[!tbp]
\includegraphics[width=0.9\linewidth]{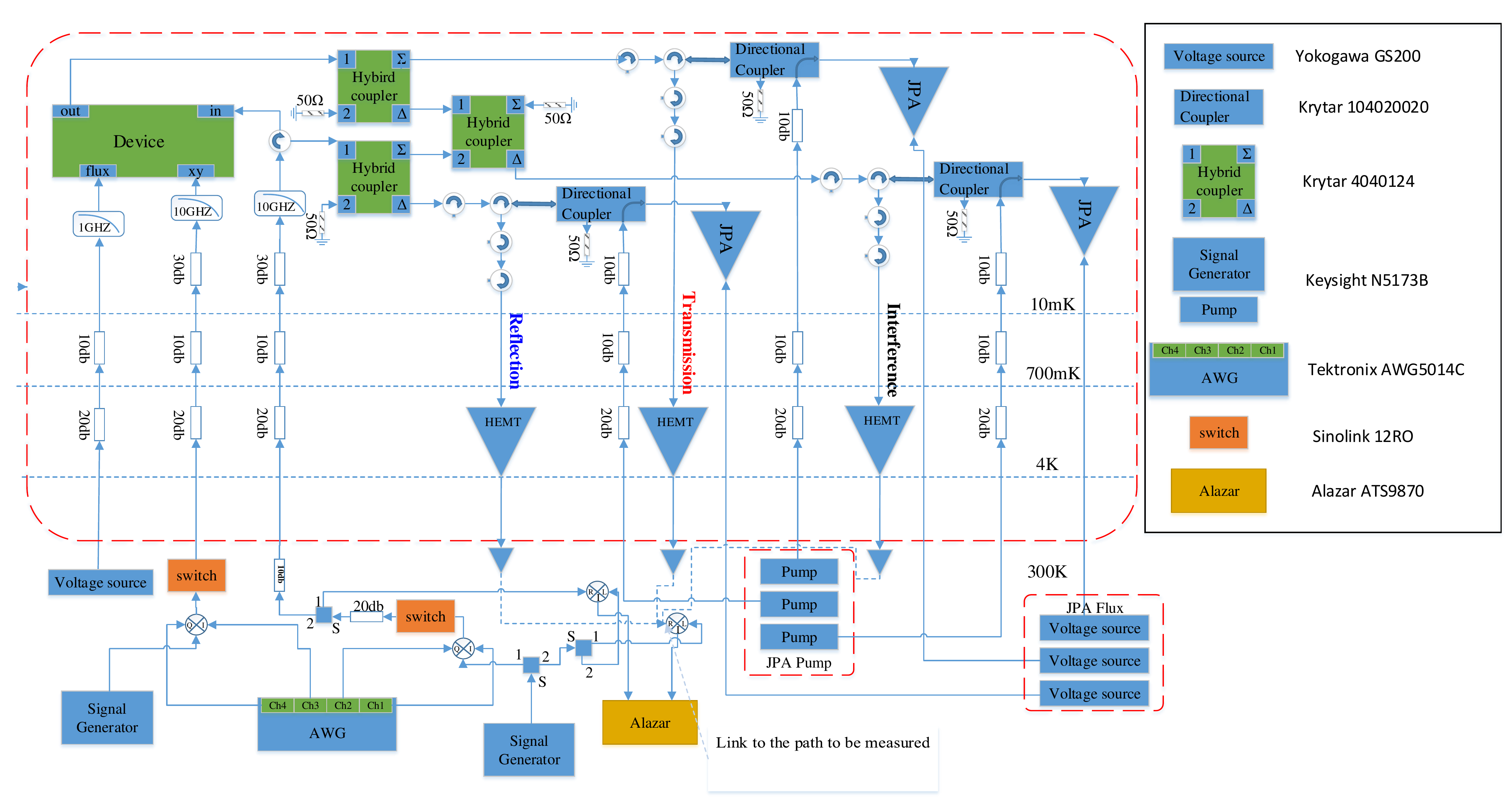}
\caption{Schematic of the experimental setup.}
\label{structure}
\end{figure*}

As mentioned in Fig.~2(b) of the main text, IQ circle of the excited state is smaller than that of the ground state, which can be explained by a finite qubit energy relaxation time $T_1$ of the qubit. The diameter of the IQ circle can be estimated based on Eq.~(\ref{ioput}) by taking $\omega_d$ as the cavity resonance or a value far off the cavity resonance. The resulting distance reads $d_g = \alpha_{in}\kappa_c/\kappa$.
For a finite $T_1$, the measured cavity response when the qubit is in $\ket{e}$ is partially mixed with that when the qubit is in $\ket{g}$, and the resulting diameter (normalized to the input state) of the IQ circle can be written as

\begin{equation}
d_e/\alpha_{in}=\kappa_c/\kappa\left|1-\exp(-\frac{t_m}{2T_1})+\exp(-\frac{t_m}{2T_1})\exp(i\delta)\right|
\label{diameter}
\end{equation}
where $\delta$ is the phase shift of the cavity response when the qubit is excited from $\ket{g}$ to  $\ket{e}$, which is about $4\chi/\kappa$. The exponential factor originates from the qubit energy relaxation. 
It is clear that $d_e$ will be smaller than $d_g$, which explains the difference between Fig.~2(b) and inset of  Fig.~2(b) in the main text.

\section{\label{sec:setup}Experimental setup}

Fig.~2(a) in the main text is a schematic of our sample and a part of the measurement setup. The sample is made from an aluminum film on a $7\,$mm$\times$7$\,$mm sapphire substrate. The Josephson junction of the transmon qubit is made by Al/AlO$_x$/Al. We use a hanger type cavity dispersively coupled to the qubit for state readout. The sample is wire bonded onto a PCB board in an aluminium sample box and cooled to about $14\,$mK by a dilution refrigerator. The measurement setup is illustrated in Fig.~\ref{structure}. Details about the device parameters are listed in Table \ref{table-1}. In particular, as shown in Fig.~\ref{structure}, three Josephson parametric amplifiers (JPA) are used in the single shot qubit state measurement, for which signal gains of $20$~dB, $14$~dB and $21$~dB for transmission, reflection and interference outputs, respectively. The JPAs are not used during the steady state spectral measurements for the cavities.

Transmission and reflection are interfered with a hybrid coupler (Krytar 4040124) as illustrated in Fig.~\ref{structure}. In order to keep the off-chip optical lengths for $T$ and $R$ as close as possible, we use three 15 inch RF cables and four SMA connectors/adapters to connect the chip with the interfering hybrid coupler for both T and R. In order to realize varied $\theta_{RT}$ for different qubits, the readout cavity for each qubit is equally spaced by about $l = 2.5\,$mm along the transmission line, which naturally serves as a path difference of $2l$ between the signals through $T$ and $R$. In the experiments we will also measure $\theta_{RT}$ for each qubit, as discussed in Appendix~\ref{sec:thetart}. It worth noting that for the purpose of applying the proposed scheme to realize simultaneous improvement for multiple cavities coupled to a common transmission line, the cavity spacing can be set to half of the wavelength corresponding to the mean resonance frequency of the cavities, and a cryogenic phase shifter is required to compensate the global relative phase difference induced by the possible imbalanced optical path between reflection and transmission.

\begin{table}
\caption{device parameters}
\begin{tabular}{cccccc}
\hline
  & Q1 & Q2 & Q3 & Q4 & Q5  \\ \hline
$\omega_r/2\pi$(GHz) & 7.9224 & 7.9756 & 8.1237 & 8.1366 & 8.1460\\
$Q_i$ & 8350&10502   & 8794 & 8391 &2580 \\
$Q_c$  & 6821& 5704 & 7044 & 3846 & 3289\\
$\omega_q$(GHz) & 5.938& 5.642  & 6.067 & 5.933 & 5.313\\
$\chi/2\pi$(MHz) & -0.4& -0.25  & -0.35 & -0.4 &  -0.35\\
$T_1(\mu s)$  &2.40 & 1.82 & 1.19 & 3.24 & 8.92 \\ 
$\theta_{RT}$ &-1.42& 0.11  &0.70 & 1.82  &2.60  \\	\hline
\end{tabular}
\label{table-1}
\end{table}

The amplification factors of the three output lines have to be carefully calibrated before making comparison among output $T$, $R$ and $T+R$. In principle it can be done by counting attenuation and amplification factors of all microwave elements on the three output lines. But in practice it would be tricky to achieve a high accuracy calibration in this way, since the calibrations have to be done at room temperature but the experiments are carried out at cryogenic temperature. Therefore, we perform the calibration by comparing spectra measured from the three outputs. 

In Fig.~2(a), after hybrid couplers the photon states from $T$, $R$ and $T+R$ can be related as
\begin{equation}
a_0^{+} = a_0^T + a_0^R
\label{seq-cali-1}
\end{equation}
As illustrated in Fig.~\ref{structure}, after passing HEMT, room temperature amplifiers and other microwave elements, the photon states get amplified and acquire an additional phase. The corresponding photon state can be written as $A^T = c^Ta_0^T$, $A^R = c^Ra_0^R$ and $A^{+} = c^{+}a_0^{+}$, where $c^T$, $c^R$ and $c^+$ represent the correspondingly amplification factors of each output line. The amplified state can be measured with a vector network analyzer. From Eq.~(\ref{seq-cali-1}) we have
\begin{equation}
A^{+} = \frac{c^{+}}{c^T}A^T +\frac{c^{+}}{c^R} A^R,
\label{seq-cali-2}
\end{equation}
which relates the three amplified state. The corresponding amplification factors can be obtained by spectra fitting according to Eq.~(\ref{seq-cali-2}), and thus the three outputs can be compared through the calibrated amplification factors.

\section{\label{sec:thetart} Calibrate the relative phase}

As indicated in Eq.~(3) in the main text, the relative phase $\theta_{RT}$ is a critical parameter to characterize the readout enhancement. $\theta_{RT}$ is related to the circuit length and microwave elements before the hybrid coupler, thus it is difficult to be determined directly. 

In this part we describe an effective approach to determine $\theta_{RT}$ for each qubit. We send a microwave probe at the frequency $\omega_r-\chi$ with an average photon number of $N$, which is on resonance with the cavity when the qubit is at the ground state $\ket{g}$. Due to the conservation of energy, we get following equations
\begin{equation}
\begin{split}
|\alpha_g^T|^2+|\alpha_g^R|^2&=r_g N,\\
|\alpha_e^T|^2+|\alpha_e^R|^2&=r_e N,
\end{split}
\label{get-gamma}
\end{equation}
where the factor 1-$r_g$ or 1-$r_e$ is the internal loss introduced by the cavity when the qubit is at $\ket{g}$ or $\ket{e}$ , which can be obtained from the spectra of the cavity. Theoretically they are given as

\begin{equation}
\begin{split}
r_g&=\left|\frac{\alpha_T(\omega_r-\chi,-1)}{\alpha_{in}}\right|^2+\left|\frac{\alpha_T(\omega_r-\chi,-1)}{\alpha_{in}}\right|^2\\
&=\frac{\kappa_c^2+\kappa_i^2}{(\kappa_c+\kappa_i)^2}\\
r_e&=\left|\frac{\alpha_T(\omega_r-\chi,1)}{\alpha_{in}}\right|^2+\left|\frac{\alpha_T(\omega_r-\chi,1)}{\alpha_{in}}\right|^2\\
&=1-\frac{2\kappa_c \kappa_i}{(\kappa_c+\kappa_i)^2+16\chi^2}
\end{split}
\label{colle-r}
\end{equation}

In order to avoid comparisons among different output lines, we focus on the same output line but different qubit states. Comparing the photon states when the qubit is at $\ket{g}$ and $\ket{e}$ for either $T$ and $R$, one would have
\begin{equation}
\begin{split}
\gamma_T &= \alpha_e^T / \alpha_g^T,\\
\gamma_R &= \alpha_g^R / \alpha_e^R
\end{split}
\label{definition-gamma}
\end{equation}
where $\gamma_T$ and $\gamma_R$ are complex numbers which contain both the amplitude and phase information. They can be measured in the experiment by comparing the output signals when the qubit is set to $\ket{g}$ and $\ket{e}$, for either of the output line $T$ or $R$.

Solving Eq.~(\ref{colle-r}) and Eq.~(\ref{definition-gamma}), the photon states can be written as
\begin{equation}
\begin{split}
\alpha_g^T&=\sqrt{N}\sqrt{\frac{r_g-r_e|\gamma_R|^2}{1-|\gamma_T\gamma_R|^2}},\\
\alpha_e^T&=\gamma_T\sqrt{N}\sqrt{\frac{r_g-r_e|\gamma_R|^2}{1-|\gamma_T\gamma_R|^2}},\\
\alpha_g^R&=\gamma_R\sqrt{N}\sqrt{\frac{r_e-r_g|\gamma_T|^2}{1-|\gamma_T\gamma_R|^2}},\\
\alpha_e^R&=\sqrt{N}\sqrt{\frac{r_e-r_g|\gamma_T|^2}{1-|\gamma_T\gamma_R|^2}}
\end{split}
\label{alphas}
\end{equation}
Two of the relative phases among these four complex amplitudes are fixed by $\gamma_T$ and $\gamma_R$, and the remaining one can be absorbed into $\theta_{RT}$ in Eq.~(\ref{alpha-int}) after interference.

Taking Eq.~(\ref{alphas}) into Eq.~(\ref{alpha-int}) we have
\begin{equation}
\begin{split}
\alpha_{g}^{+}&=\frac{\alpha_{g}^T+e^{i\theta_{RT}}\alpha_{g}^R}{\sqrt{2}}\\
&=\sqrt{\frac{N}{2}} \left( \sqrt{\frac{r_g-r_e|\gamma_R|^2}{1-|\gamma_T\gamma_R|^2}} + \gamma_R e^{i\theta_{RT}}  \sqrt{\frac{r_e-r_g|\gamma_T|^2}{1-|\gamma_T\gamma_R|^2}}\right),\\
\alpha_{e}^{+}&=\frac{\alpha_{e}^T+e^{i\theta_{RT}}\alpha_{e}^R}{\sqrt{2}}\\
&=\sqrt{\frac{N}{2}} \left( \gamma_T\sqrt{\frac{r_g-r_e|\gamma_R|^2}{1-|\gamma_T\gamma_R|^2}} + e^{i\theta_{RT}}\sqrt{\frac{r_e-r_g|\gamma_T|^2}{1-|\gamma_T\gamma_R|^2}}\right).
\end{split}
\label{interference-photon}
\end{equation}
The amplitude ratio of the interference signal $\gamma_+=\alpha_{e}^+/\alpha_{g}^+$ can also be measured in the experiment. The expression of $\theta_{RT}$ can be deduced based on Eq.~(\ref{alpha-int}) and Eq.~(\ref{alphas}), which reads as
\begin{equation}
\exp(i\theta_{RT})=\frac{\alpha_e^{T}-\gamma_{+}\alpha_g^{T}}{\gamma_{+}\alpha_g^{R}-\alpha_e^{R}} =\frac{(-\gamma_{+} + \gamma_T) \sqrt{r_g-r_e|\gamma_R|^2}}{(-1+ \gamma_{+} \gamma_R) \sqrt{r_e-r_g|\gamma_T|^2}}.
\label{rtangle}
\end{equation}

In the experiment, we first set the qubit in either $\ket{g}$ or $\ket{e}$, then send the cavity probe signal at frequency $\omega_r-\chi$ with certain average photon numbers. On the output line of $T$, $R$ and $T+R$, the photon states $\alpha_{0(1)}^T$, $\alpha_{0(1)}^R$ and $\alpha_{0(1)}^+$ can be measured, from which $\gamma_R$ , $\gamma_T$ and $\gamma_+$ can be deduced from their definitions. All of the measurements are repeated for 40 times to reduce statistical errors. Then the value of $\theta_{RT}$ can be calculated based on Eq.~(\ref{rtangle}). The results are shown in the inset of Fig.~3 of the main text as the horizontal axis.

\section{\label{sec:singleshot}Single shot measurement}

In the single shot experiment we compare histogram plots of the measured cavity response and the resulting fidelity when measuring from output $T$ and output $T+R$. Since the amplification and detection efficiency of the two output circuits are not necessarily the same, we have to first calibrate the circuit parameters before the comparison.
In the measurement setup we use homodyne detection to map the amplified output photon state to the voltages, therefore  $\langle \alpha \rangle$ and the variance of measured photon state is mapped as the mean value and the variance of the measured voltage distribution. For a coherent state input readout signal, the output signal can be regarded as a coherent state convoluted with the added noise from the amplification chain.
This added noise does not change the mean value of the measured voltage, but it broadens the voltage distribution and thus leads to a larger variance $\sigma_m$ \cite{Eichlerthesis}. Here we can define the circuit efficiency $\eta=1/(1+N_0)$ to describe the added noise $N_0$.
It has been shown that the noisy variance $\sigma_m$ and the original variance $\sigma_0$ are related as $\sigma_m=\sigma_0/\sqrt{\eta}$ \cite{Martinis15}.

\begin{figure}[!tbp]
\includegraphics[width=1\linewidth]{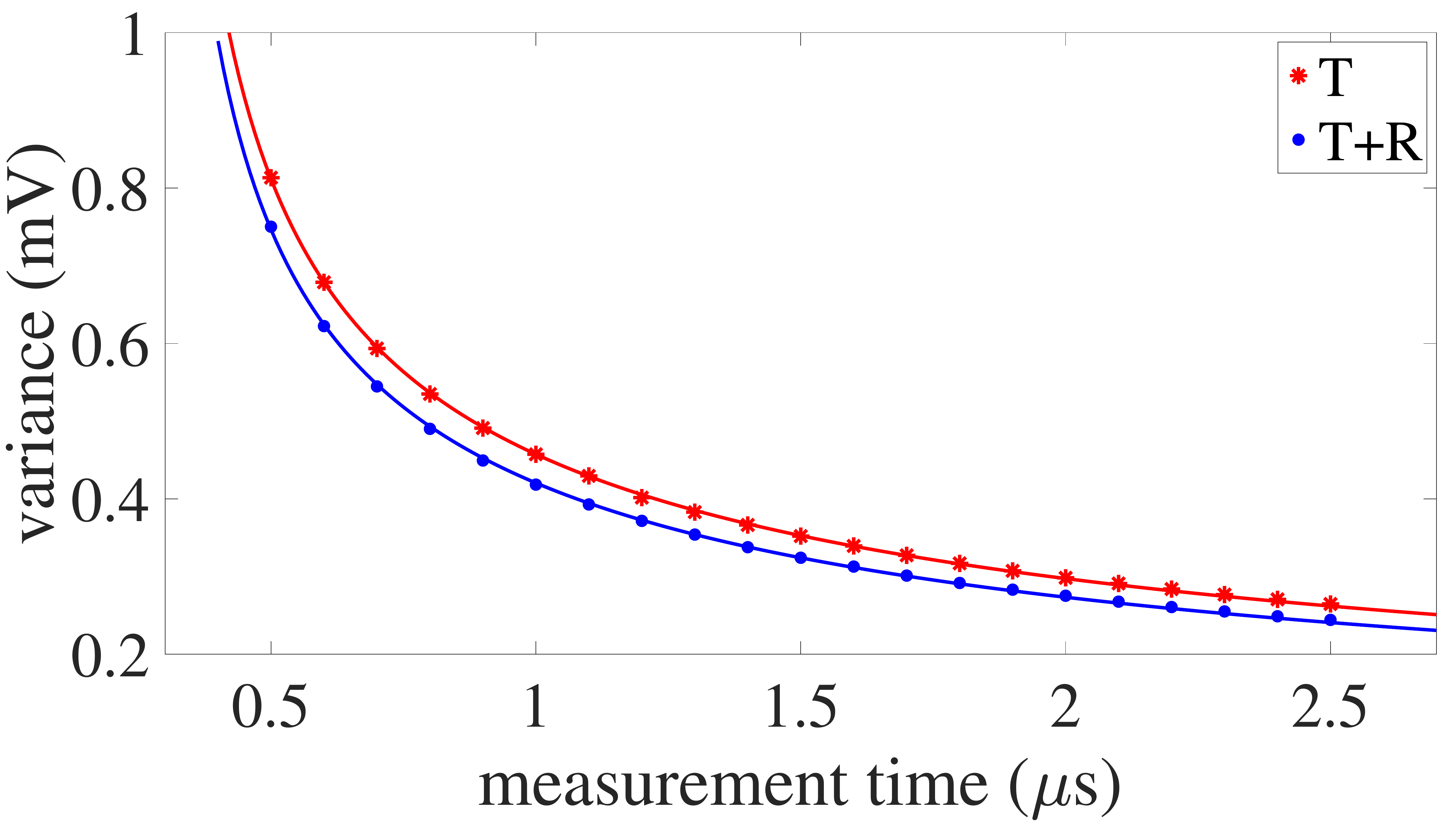}
\caption{The variance of the readout signal as a function of measurement time for different circuits. The blue (red) scattered plots are experimental data from output $T$ ($T+R$), and the lines are the corresponding fitting results.}
\label{eta}
\end{figure}

The circuit efficiency can be calibrated by measuring the variance of voltage distribution with different measurement time $t_m$. Using the fitting model $\sigma_m=c_0/\sqrt{t_m}$, we can get the parameter $c_0$ which is proportional to $1/\sqrt{\eta}$  \cite{Martinis15}.
The corresponding measurement data and fitting results for output $T$ circuit and $T+R$ circuit are shown in Fig.~\ref{eta}.
We get $c_0^{T+R}=0.3598\pm0.0008$ and $c_0^T=0.3915\pm0.0006$ respectively, with the ratio of the circuit efficiencies $\eta_{T+R}/\eta_T=(c_0^T/c_0^{T+R})^2=1.1838$.
When comparing the measurement results between $T$ and $T+R$, 
the variance of the histogram result for $T+R$ output $\sigma_{T+R}$ is rescaled as $\sigma_{T+R}'=c_0^T/c_0^{T+R}\sigma_{T+R}$ to remove the efficiency difference between two output circuits. After this process, Fig.~4(c) in the main text can be obtained.

Now we discuss the potential of reducing the total readout error when using path interference compared with using simply transmission or reflection. As mentioned in the main text, the total readout error can be written as the sum of measurement error, qubit relaxation induced error and thermal population induced error, 
The thermal population induced error $P_{th}$ can be estimated via the Boltzmann distribution, $\frac{\exp(-\hbar\omega_q/k_bT_{e})}{1+\exp(-\hbar\omega_q/k_bT_{e})}$~\cite{Delsing16}, where $T_e$ is the device temperature. The thermal population induced error can be greatly suppressed by improving the thermal anchoring of the sample and coaxial cables and using carefully selected filters and attenuators~\cite{Wallrafffridge2019}, which we will not discuss here.

The measurement error $P_m$ can be expressed in terms of measurement time $t_m$ and distance $D$ between two pointer states corresponding to the two qubit states, $P_m = 1-\mbox{Erf}(\sqrt{\eta t_m}D/\sqrt{2})$ \cite{Martinis15}. The distance of the pointer states $D$ is related to the cavity photon number $n_c$ as $D^T=\frac{\sqrt{2\kappa_c n_c}}{\sqrt{1+(\frac{\kappa}{2\chi})^2}}$, where we take the distance on the output $T$ $D_T$ as an example.
Neglecting the thermal population induced error, we have 
\begin{equation}
\begin{split}
P_{err} &\sim P_m+P_{T_1} \\
&=1-\exp(-\frac{t_m}{2 T_1})+1-\mbox{Erf}(\sqrt{\eta t_m}D/\sqrt{2}) .\\
\label{err-fid-apd}
\end{split}
\end{equation}
From Eq.~(\ref{err-fid-apd}), a longer measurement time is preferred to reduce the measurement error, but on the other hand will increase the qubit relaxation induced error.
We can define an optimized measurement time $t_{optimal}$ which yields the minimum total readout error. By taking the derivative of Eq.~(\ref{err-fid-apd}),  $t_{optimal}$ can be obtained as
\begin{equation}
t_{optimal}=\frac{T_1}{\eta D^2T_1-1}\mbox{W}(\frac{2\eta D^2T_1}{\pi}(\eta D^2T_1-1)), 
\label{t_min}
\end{equation}
where W is Lambert W function. The minimum qubit readout error $P_{err}^{min}(T_1,D)$ can be obtained by plug $t_{optimal}$ back into Eq.~(\ref{err-fid-apd}). As a specific example, we take the circuit efficiency $\eta=0.25$, the cavity photon number $n_c=20$, the device temperature $T_e \sim 20\,$mK, and the qubit frequency $\omega_q \sim 6\,$GHz based on our measurement circuit. Then the total readout error with the optimized measurement time for $T$ and $T+R$ outputs can be obtained according to Eq.~(\ref{err-fid-apd}) and Eq.~(\ref{t_min}). The corresponding results are shown in Fig. 4(d) of the main text.

\bibliography{ref}